\documentclass[twocolumn]{aastex631}
\received{30 Aug 2021} 
\accepted{13 Dec 2021 in The Astronomical Journal}

\begin{document}

\title{On spectroscopic phase-curve retrievals: H$_2$ dissociation and thermal inversion in the atmosphere of the ultra-hot Jupiter WASP-103\,b.}

\author[0000-0001-6516-4493]{Quentin Changeat}
\affil{Department of Physics and Astronomy, University College London, London, United Kingdom}

\begin{abstract}
    This work presents a re-analysis of the spectroscopic phase-curve observations of the ultra hot Jupiter WASP-103\,b obtained by the \emph{Hubble Space Telescope} (\emph{HST}) and the \emph{Spitzer Telescope}. Traditional 1D and unified 1.5D spectral retrieval techniques are employed, allowing to map the thermal structure and the abundances of trace gases in this planet as a function of longitude. On the day-side, the atmosphere is found to have a strong thermal inversion, with indications of thermal dissociation traced by continuum H$^-$ opacity. Water vapor is found across the entire atmosphere but with depleted abundances of around 10$^{-5}$, consistent with the thermal dissociation of this molecule. Regarding metal oxide and hydrides, FeH is detected on the hot-spot and the day-side of WASP-103\,b, but TiO and VO are not present in detectable quantities. Carbon-bearing species such as CO and CH$_4$ are also found, but since their detection is reliant on the combination of \emph{HST} and \emph{Spizer}, the retrieved abundances should be interpreted with caution. Free and Equilibrium chemistry retrievals are overall consistent, allowing to recover robust constraints on the metallicity and C/O ratio for this planet. The analyzed phase-curve data indicates that the atmosphere of WASP-103\,b is consistent with solar elemental ratios. \vspace{15mm}
\end{abstract}

\section{Introduction}

In the last decade, the study of exoplanets was revolutionized by the launch of the first dedicated missions, the \emph{Convection, Rotation et Transits planétaires Space Telescope} \cite[\emph{CoRoT}: ][]{CoRoT_2009} and the \emph{Kepler Space Telescope} \cite[\emph{Kepler}: ][]{borucki_kepler}. These telescopes, along with the recent \emph{Transiting Exoplanet Survey Satellite} \cite[\emph{TESS}: ][]{TESS_2015_ricker}, utilized the transit method to discover more than 4000 worlds and provided crucial information regarding their radius. Combined with complementary constraints on exoplanet orbital properties and masses, obtained by radial velocity measurements from the ground, a first picture of the exoplanet population emerged. Building from the discoveries of these missions, exoplanets are now known to be extremely diverse. While small rocky worlds are clearly more frequent \citep{Howard_2010,Burke_2015, Fulton_2017}, biases induced by our observing method have revealed many objects that have no obvious counterparts in the solar system, such as hot Jupiters and transitional worlds. The existence of such objects came as a surprise and challenged our understanding of planetary formation and evolution. 

The proximity of hot Jupiters to their host stars suggests that significant migration must have happened during their formation but the details of these processes remain unknown \citep{Ikoma_2000, Udry_2003, papaloizou_2006, Ford_2008, Bailey_2018, Dawson_2018, Shibata_2020}. Since exoplanets capture their materials from protoplanetary disks, which have an in-homogeneous composition in gas and solids, one way these processes could be investigated is by observing the bulk composition of exoplanets \citep{madhu_formation, Mordasini_2016, Eistrup_2018, Lothringer_2020, Turrini_2021, Hobbs_2021}. Similarly, transitional planets, with radii between 1.2 Re and 2.5 Re, are still poorly understood. The nature of these planets cannot be determined from their radius and mass alone due to the degeneracies in their internal composition \citep{Adam_2008, Valencia_2013, Hu_2015, Kimura_2020, Madhu_2020_k218, changeat_2020_disentangling, Kite_2021}. From considerations of their density, smaller planets are thought to be rocky with a metallic core and a large rocky mantel, while larger planets might have a core but are mostly made of hydrogen and helium. For transitional planets, multiple regimes combining hydrogen, water, rocks, and denser elements can exist and as of today, it is not known whether the transition between sub-Neptunes and super-Earth is sharp or smooth. Radii and masses only provide a limited window into exoplanets, and in order to further understand the observed population, the field is now moving towards the characterization of their atmospheres.

The characterization of exoplanet atmospheres is typically done via spectroscopic observations of transits and eclipses, when the planet passes respectively in front or behind its host star. These observations, often carried by the \emph{Hubble Space Telescope} (\emph{HST}) and the \emph{Spitzer Space Telescope} (\emph{Spitzer}),  allowed to constrain the day-side and the terminator of many targets. Using both transit and eclipse from space and the ground, ultra-hot Jupiters \citep{swain_wasp12, Haynes_Wasp33b_spectrum_em, Evans_wasp121_e1, Evans_wasp121_e2, VonEssen_2019, Evans_wasp121_e3, Edwards_2020_ares, Ehrenreich_2020, pluriel_aresIII, hoeij_k9, hoeij_k9_2, Tabernero_2021, Changeat_2021_k9}, hot Jupiters \citep{charbonneau_2002, mccull_hd189, Crouzet_HD189_spectrum_em, swain_2008_hd189_nicmos, swain_hd189_nicmos_tr, Swain_2008_hd209, Swain_2009_hd209, tinetti_water, Tinetti_2010, Line_HD209_spectrum_em, Cabot_2019,skaf_aresII,  Anisman_2020, changeat_kelt11, Yip_2021_w96, Giacobbe_2021_5species, Dang_2021} but also smaller Neptunes-size objects \citep{Beaulieu_2011, Kulow_2014, Allart_2018, Palle_2020, Guilluy_2021}, transitional planets \citep{Kreidberg_GJ1214b_clouds, Tsiaras_2019_k218, Benneke_2019_k218} and even rocky planets \citep{tsiaras_55cnce, Dewit_2016_trap,  Demory_2016, DeWit_2018_trap, kreidberg_lhs, Edwards_2021_lhs,  Mugnai_2021, Gressier_2021} were probed. Except for the smallest targets, which remain very challenging even to this date, the use of transit and eclipse spectroscopy in exoplanets is now routine and population studies of exo-atmospheres \citep{sing_pop, tsiaras_30planets, pinhas, Welbanks_2019} have pushed our understanding of these worlds much beyond the mass-radius picture. 

There are however major limitations to observations in transit and eclipse spectroscopy. Typically, these methods only provide 1-dimensional (1D) constraints on atmospheric properties that are inherently 3-dimensional (3D). The 3D nature of tidally locked exoplanets has been highlighted many times in the literature, principally via Global Climate and Atmospheric Dynamics Models \citep{Cho_2003, Showman_2010, Leconte_2013, mendonca_2016, Parmentier_2018_w121photodiss, Showman_2020, Skinner_2021, Cho_2021}. In addition, due to the lack of 3D constraints from transit and eclipse observations, models often assume 1-dimensional geometries that can lead to important biases in the recovery of chemical abundances or temperatures \citep{Feng_2016, Caldas_2019, Changeat_2020_pc1, Pluriel_2020_3d, Taylor_2020, Macdonald_2020, Espinoza_2021}. With the upcoming launch of next-generation telescopes such as the \emph{James Webb Space Telescope} \cite[\emph{JWST}: ][]{Greene_2016} and \emph{Ariel} \citep{Tinetti_ariel, Tinetti_2021}, which will be much more sensitive to 3D effects, overcoming these issues is of major importance.

Spectroscopic phase-curve observations provide a natural solution to beak the degeneracies arising from the 3D aspects of exoplanets. In phase-curve, the planet's emission is tracked along a full orbit, allowing to recover the atmospheric information as a function of longitude. Phase-curve observations from space have so far been mainly obtained by \emph{TESS} (0.8$\mu$m channel), \emph{HST} (G141 grism from 1.1$\mu$m to 1.6$\mu$m) and \emph{Spitzer} (3.6$\mu$m and 4.5$\mu$m channels), but due to the much longer time and the more complex characteristics of those observations, only a handful of planets have been characterized using this technique. The most notable exoplanets that were observed with \emph{HST} and \emph{Spitzer} so far are WASP-43\,b \citep{Stevenson_2014_w43, Stevenson_2017_w43}, WASP-103\,b \citep{Kreidberg_w103}, WASP-18\,b \citep{arcangeli_w18_phase}, WASP-12\,b \citep{Arcangeli_2021_12pc} and WASP-121\,b \citep{Evans_wasp121_PC}. Phase-curves obtained with \emph{Spitzer} only are more common but since only two photometric channels are obtained, deep characterization of exo-atmospheres with those datasets is more limited. Additionally, the continuous and extended sky coverage of TESS has allowed to obtain a multitude of phase-curve observations in the visible \citep{shporer_w18, wong_k9, wong_wasp19_tess, wong_2020_tess1, wong_2021_tess2, bourrier, von_essen_2020_tess, Jansen_2020, Beatty_2020_tess}. These photometric observations provide complementary constraints on planetary albedo, thermal redistribution, and clouds due to their lower wavelength ranges.

Since phase-curve observations provide 3-dimensional constraints, their analysis requires more complex and computing-heavy models. Often, forward model comparisons, using complex 3-dimensional atmospheres, are used. However, the lack of statistical frameworks and the strong assumptions on the chemical, thermal and dynamical processes they require, implies that the parameter space of possible solutions for a given dataset is poorly explored. Atmospheric retrievals, which are the most popular statistical tools to extract atmospheric properties from exoplanet transit and eclipse data, can also be used in the context of phase-curve \citep{Stevenson_2017_w43, Changeat_2020_pc1,Irwin_2020, Changeat_2020_pc2, Feng_2020}. Early works used traditional 1D retrievals at different phases to extract the 3D dependence of thermal profiles, chemical and cloud properties \cite[e.g: ][]{Stevenson_2017_w43}. More recent studies have used unified retrievals, which can simulate the phase-dependent emission, fitting all the data at once, and thus properly exploiting redundant information between the different phases. Such models were developed and employed to study the exoplanet WASP-43\,b revealing a much more complex atmosphere than previously thought \citep{Changeat_2020_pc2}. This new work investigates the atmosphere of the ultra-hot Jupiter WASP-103\,b using traditional 1D and unified 1.5D retrieval techniques, demonstrating the added benefits of phase-curve observations as compared to the more common transit and eclipse observations. Understanding the processes driving atmospheric observations of close-in planets is crucial for ensuring an unbiased link between atmospheric properties and planetary formation. 

WASP-103\,b \citep{gillon_wasp103} is an ultra-hot Jupiter that orbits its host star, an F8V star with a temperature of 6600K, in 22 hours. The large planet (1.528 R$_J$) is ideal for spectroscopic observations in transit, eclipse, and phase-curves due to its high equilibrium temperature of about 2500K. Observations of the phase-curve emission of WASP-103\,b, obtained by \emph{HST} and \emph{Spitzer}, were reported in \cite{Kreidberg_w103}. In their study, they found a large day-night contrast and a symmetric emission along the entire orbit, which suggest a poor energy redistribution from the day-side to the night-side. Via retrievals, their study highlighted the presence of a day-side thermal inversion, likely associated with a strong optical absorber (TiO, VO, or FeH). Their study did not detect water vapor, which is present in most hot Jupiters. This lack of detection was attributed to thermal dissociation processes, which are believed to be important in ultra hot Jupiters \citep{Lothringer_2018, Parmentier_2018_w121photodiss, gandhi_w18, Changeat_2021_k9}.
Complementary studies, using observations from the ground with \emph{Gemini/GMOS} \citep{Wilson_2020_W103}, \emph{VLT/FORS2} \citep{lendl_wasp103}, the \emph{ACCESS} survey and the \emph{LRG-BEASTS} survey \citep{Kirk_2021_w103}, presented contradictory evidence for the presence of water vapor. These studies also indicated the possible presence of other absorbers such as K, Na, HCN and TiO. 

Here, the phase-curve observations reduced by \cite{Kreidberg_w103} are re-analyzed using the suite of retrieval tools \emph{TauREx3} \citep{2019_al-refaie_taurex3, al-refaie_2021_taurex3.1}. As compared to \cite{Kreidberg_w103}, which only performed an MCMC grid-based atmospheric model fitting of the eclipse and the transit, here, the entire phase-curve data is considered in the atmospheric retrievals. The analysis is performed using a wide suite of atmospheric models, including 1D and 1.5D models, and a robust nested sampling optimization technique.

\section{Methodology}

The observations of WASP-103\,b are taken as-is from the previous work of \cite{Kreidberg_w103}. These consist in the reduced spectra for the \emph{HST} Wield Field Camera 3 (WFC3), Grism G141, covering the range 1.1$\mu$m to 1.6$\mu$m, and the \emph{Spitzer} 3.6$\mu$m and 4.5$\mu$m channels. The observations span the entire orbit in normalized phase steps of 0.1 from 0.0 to 0.9, with phase 0.0 corresponding to the transit and phase 0.5 corresponding to the eclipse. The observations are already corrected for the faint background star \citep{Wollert_2015, Ngo_2016}. In discussion, comparisons of the transit observations from \cite{Kirk_2021_w103} with \emph{Gemini/GMOS}, \emph{VLT/FORS2}, \emph{ACCESS} and \emph{LRG-BEASTS}, against the \emph{HST} and \emph{Spitzer} phase-curve retrievals are also presented. 

To analyze the phase-curve observations of WASP-103\,b, the atmospheric framework \emph{TauREx3}\footnote{ \emph{TauREx3.1} is used: \url{https://github.com/ucl-exoplanets/TauREx3_public}}. \emph{TauREx3} is a library dedicated to the study of exoplanetary atmospheres. It includes a large selection of radiative transfer models and chemistries, as well as state-of-the-art optimization techniques for performing atmospheric retrievals. In its last version, \emph{TauREx3.1} \citep{al-refaie_2021_taurex3.1}, the concept of plugins was introduced. Plugins are separated plug-and-play codes that enhance the functionalities of \emph{TauREx} without requiring modifications of the main code. This flexibility makes it easy to add new features and models such as the \emph{GGChem}\footnote{ \emph{GGChem} plugin for \emph{TauREx}: \url{https://github.com/ucl-exoplanets/GGchem}} chemistry code \citep{Woitke_2018} or the 1.5D phase-curve model \citep{Changeat_2020_pc1, Changeat_2020_pc2} that are central to this study. 

In this work, two retrieval techniques implemented in \emph{TauREx3} and employing GPU acceleration are utilized. First, a traditional 1D retrieval is performed for each phase spectrum. Second, a 1.5D unified retrieval is used to fit all the phases at once. For each technique, multiple scenarios are presented. In each scenario, the parameter space is explored using the \emph{MultiNest} optimizer \citep{Feroz_multinest, buchner_multinest} with an evidence tolerance of 0.5 and 500 live points. The nested sampling optimization being handled with \emph{MultiNest}, the Bayesian Evidence of each retrieval is automatically obtained and used for model comparison \citep{Kass1995bayes, jeffreys1998theory, Knuth_2014, Robert_2008, Waldmann_taurex1}. For each run, an error estimate of the Bayesian Evidence is also reported \citep{Feroz_hobson_2008}. Details on the retrieval setups can be found bellow.

\subsection{1D individual retrievals}

Individual 1D retrievals (1D) are performed for each phase-curve spectra using a similar technique to what is presented in \cite{Stevenson_2017_w43}. For each scenario, this constitutes 10 retrievals from phases 0.0 (transit) to 0.9, with phase 0.5 being the eclipse. The \emph{TauREx} emission retrievals assume a 1D parallel atmosphere with 100 layers, equally space in log space from 10$^6$ Pa to 0.1 Pa. The atmosphere is considered primary and filled with hydrogen and helium at solar ratio. Molecular line-lists are taken from the Exomol project \citep{Tennyson_exomol, Chubb_2021_exomol}, HITEMP \citep{rothman} and HITRAN \citep{gordon}. Opacities for the trace gases H$_2$O \citep{barton_h2o,polyansky_h2o}, CH$_4$ \citep{hill_xsec,exomol_ch4}, CO \citep{li_co_2015}, CO$_2$ \citep{Yurchenko_2020}, TiO \citep{McKemmish_TiO_new}, VO \citep{McKemmish_2016_vo} and FeH \citep{Bernath_2020_FeH}, as well as continium H$^-$ opacity \citep{John_1988_hmin,Edwards_2020_ares}, are considered in this work. In addition, Rayleigh scattering \citep{cox_allen_rayleigh} and Collision Induced Absorption (CIA) from H$_2$-H$_2$ and H$_2$-He pairs are considered \citep{abel_h2-h2,fletcher_h2-h2,abel_h2-he}. Given the planet's equilibrium temperature \cite[T=2508K,][]{gillon_wasp103} and the results from previous studies, cloud/hazes opacities are not considered in this work.

For the thermal profile, an NPoint parametric model, interpolating between N temperature-pressure nodes is employed. The scenario presented for the 1D individual runs possesses 7 nodes (labeled 7PT) at fixed pressures. The pressure points are fixed at 10$^6$, 10$^5$, 10$^4$, 1000, 100, 10, and 0.1 Pa. A detailed investigation on free thermal profile parametrizations can be found in Appendix D of \cite{Changeat_2020_pc2}. The type of setup used here was found relevant and practical for phase-curve data with 1.5D models. A more conservative model with 3 nodes (labeled 3PT) was also tested. In this case, both pressure and temperature are retrieved, with retrieval bounds from 10$^6$ Pa to 0.1 Pa for the pressure and 300K to 5000K for the temperature. The 3PT case obtained similar results and Bayesian Evidence to the 7PT case due to the retrieved thermal profile being well described by three freely moving nodes, so only the 7PT results are presented here.

Considering the chemistry, free and equilibrium chemical models are tested. In the free runs (FREE), the abundance of each species, modeled as constant with altitude, is retrieved individually. For H$^-$, the abundance of H is fixed to a pre-defined profile \cite[similar to ][]{Edwards_2020_ares}, while only e$^-$ is retrieved. In the equilibrium case (EQ), the chemical code \emph{GGChem} \citep{Woitke_2018}, which recently received a \emph{TauREx3} plugin, is employed. Assuming equilibrium chemistry, only the metallicity and the C/O ratio is retrieved.

In the emission spectra, the planetary radius and mass are not retrieved due to the availability of more precise constraints obtained in transit and Radial Velocity surveys \citep{changeat_mass}. 

\subsection{1.5D unified retrievals}

In the unified retrievals (1.5D), the planet is separated into three regions: hot-spot, day-side, and night-side. The retrieval attempts to fit all the observed spectra at once, including transit, recovering the best set of parameters for each region. A complete description of the model, which was previously used on the WASP-43\,b phase-curve can be found in \cite{Changeat_2020_pc1, Changeat_2020_pc2}.

Due to performance considerations, it is impractical to retrieve the abundance of each species independently in each region, which would lead to 24 free parameters for the chemistry alone. The observations are therefore first fitted using the equilibrium setup (EQ) and temperature profiles using 7 fixed nodes for each region. Then for the free run, the temperature profile is fixed to the best-fit profile of the EQ runs, which greatly reduces the dimension of the retrieval. A 3 node temperature-pressure profile case was also run for comparison. This case obtained a lower Bayesian evidence (885.8$\pm$0.3) than the 7PT case presented in the result section (892.3$\pm$0.3).

The geometry of the phase-curve model requires information regarding the size of the hot-spot region and its offset. For the hot-spot offset, it is fixed to 0.0 degrees, following the findings from \cite{Kreidberg_w103}. For the hot-spot size, 40, 60, 70, and 80 degrees were tested. The 60, 70, and 80 degree cases obtained similar Bayesian evidence, respectively 890.3$\pm$0.3, 892.3$\pm$0.3 and 892.2$\pm$0.3, while the 40 degree case obtained only 886.4$\pm$0.3, which suggest a large hot region on the day-side and a sharp day-night transition. In the result section, only the 70 degree case is presented as it better matches the 1D results and the findings from \cite{Kreidberg_w103} and that the conclusions are the same with the 60 and 80 degree cases.

\section{Results}

The results of the 1D and 1.5D retrievals are detailed in the next sections. For the transit fits, which are included in the 1.5D retrievals, the conclusions are discussed in the Discussion section.

\subsection{1D retrievals}

\begin{figure}
    \includegraphics[width = 0.5\textwidth]{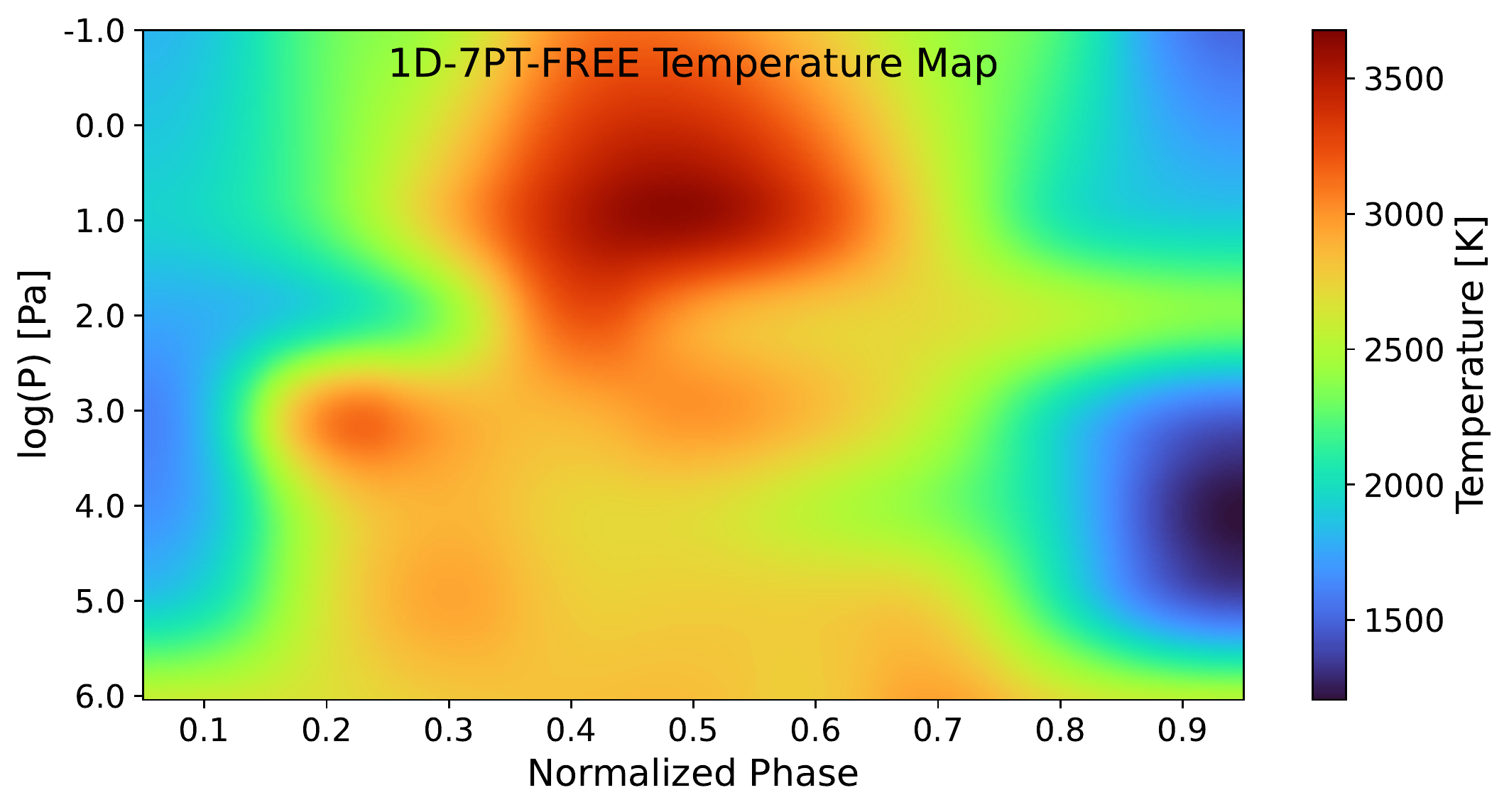}
    \includegraphics[width = 0.5\textwidth]{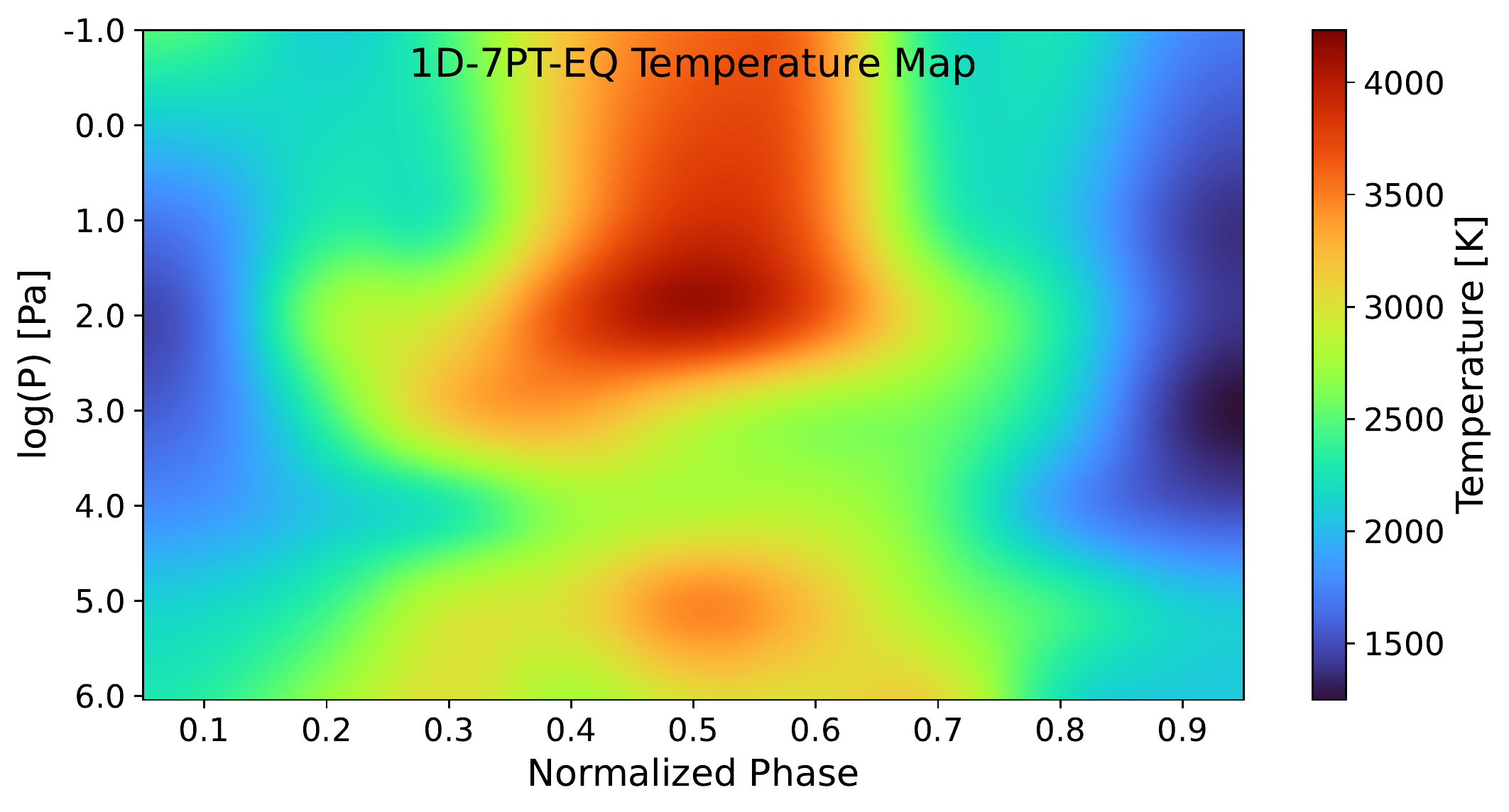}
    \caption{Recovered mean thermal structure from the 1D-7PT retrievals. Top: FREE scenario; Bottom: EQ scenario.}
    \label{fig:tp_map_free}
\end{figure}

For the 1D retrievals, 7PT models provide a similar to slightly higher Bayesian Evidence to 3PT models. The added complexity might be debatable but since the results are very similar in both cases, and for consistency with the 1.5D results, the 7 node retrievals are shown here. The observations and best-fit spectra obtained using the 1D retrievals are presented in Appendix 1.  In Figure \ref{fig:tp_map_free}, the recovered thermal structure is shown for both the FREE and EQ cases.

\begin{figure}
    \includegraphics[width = 0.5\textwidth]{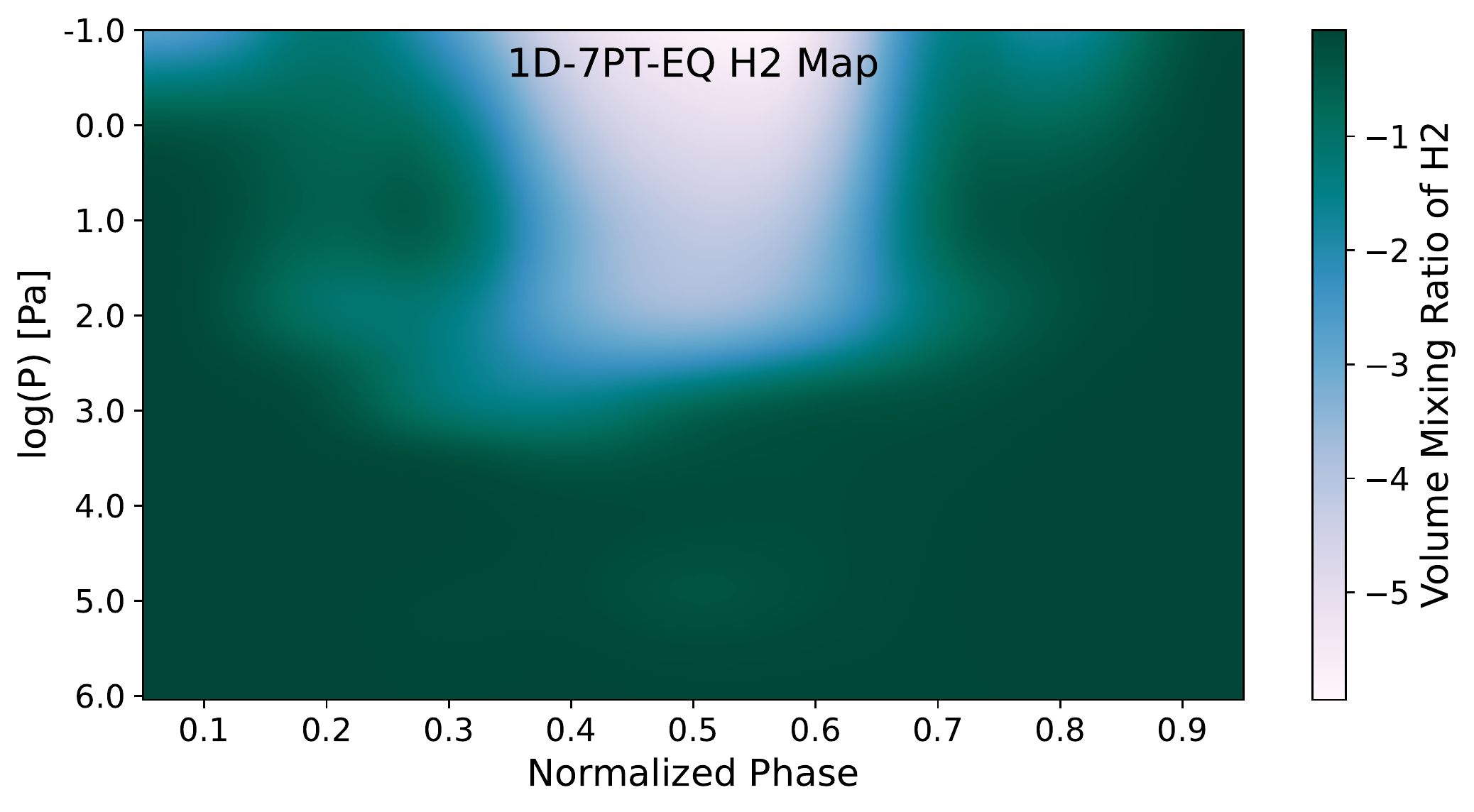}
    \includegraphics[width = 0.5\textwidth]{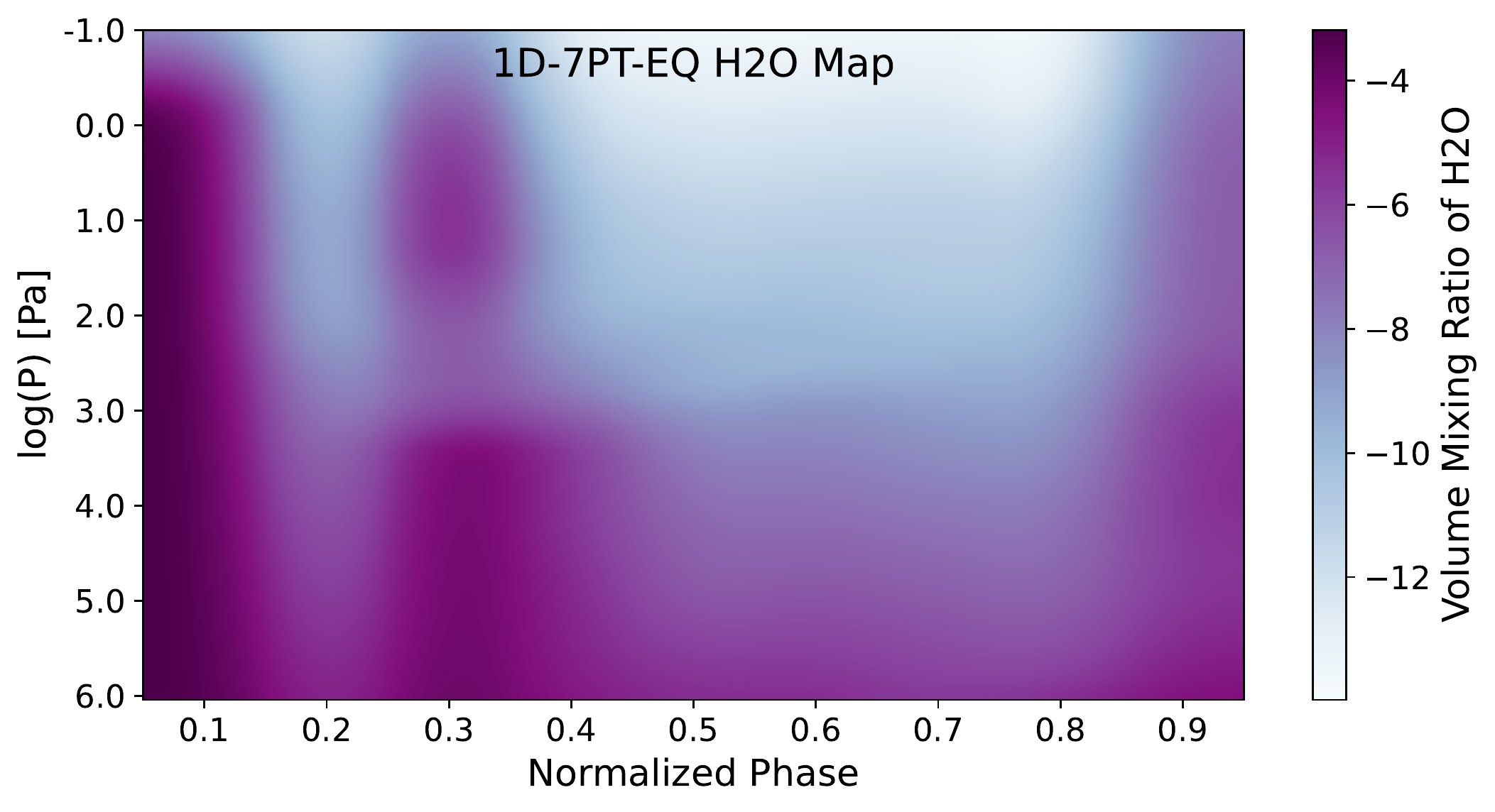}
    \caption{Recovered mean abundances from the 1D-7PT-EQ retrieval. Top: H$_2$; Bottom: H$_2$O. Those maps highlight the thermal dissociation processes at the highest altitude of WASP-103\,b day-side.}
    \label{fig:h2o_h2_map_free}
\end{figure}

The atmosphere exhibit a strong, extended thermal inversion on the day-side, which is consistent with the findings from \cite{Kreidberg_w103}. The large day-night temperature contrast is indicative of a poor energy redistribution. Similar maps can also be obtained for the chemical species in the EQ scenario. In Figure \ref{fig:h2o_h2_map_free}, such map is shown for the H$_2$O distribution. The H$_2$O abundance distribution, while obtained under the assumption of equilibrium chemistry, highlights the presence of thermal dissociation processes on the day-side of WASP-103\,b, in the highest part of its atmosphere. This is associated with an increase in neutral H, which becomes the dominant species, and e$^-$. The abundance of e$^-$ directly informs on the strength of bound-free and free-free absorption from H$^-$ \citep{john_1988_h-}, which leads to continuum absorption and could explain the relatively featureless spectra in the shorter wavelengths of WFC3. Given the very high retrieved temperatures of WASP-103\,b, similar processes must also happen for the main metal oxides and hydrides TiO, VO, and FeH, as seen in the maps of those species that are available in Appendix 2.

\begin{figure}
    \includegraphics[width = 0.5\textwidth]{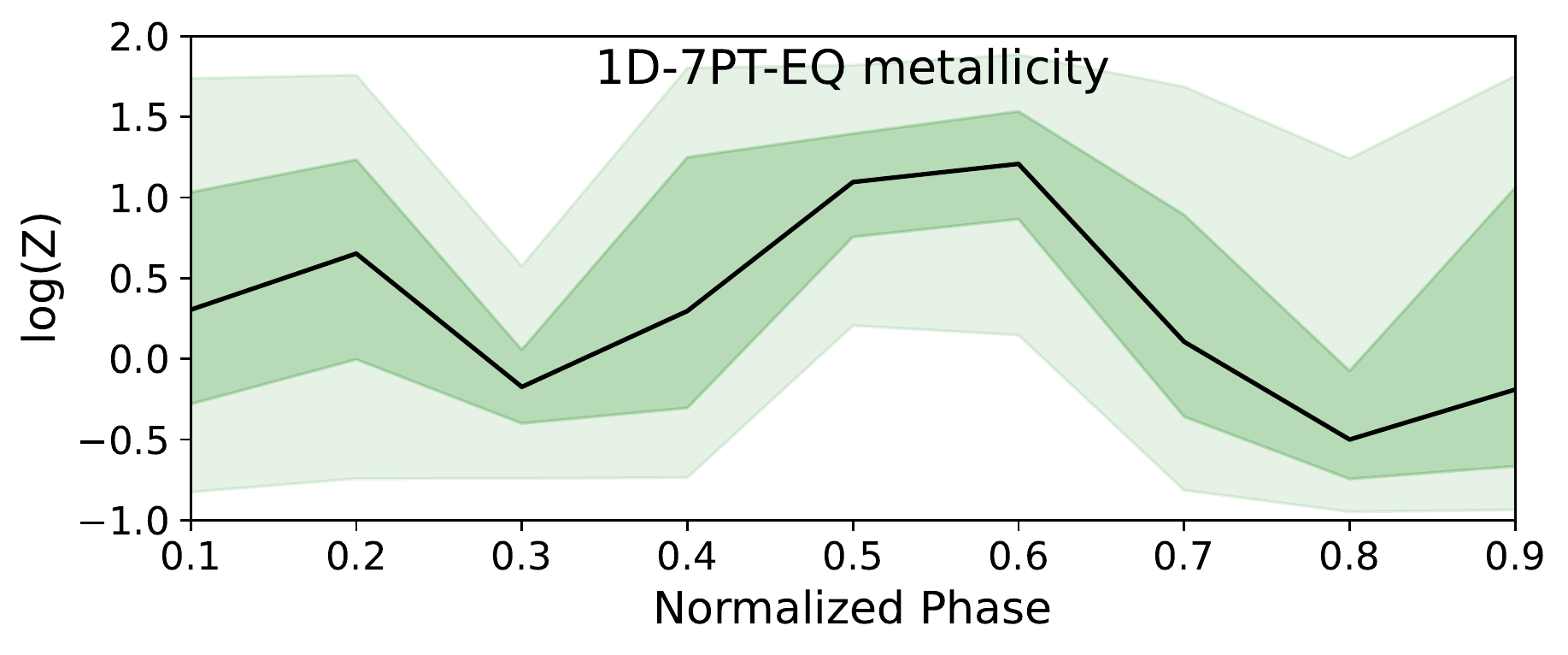}
    \includegraphics[width = 0.5\textwidth]{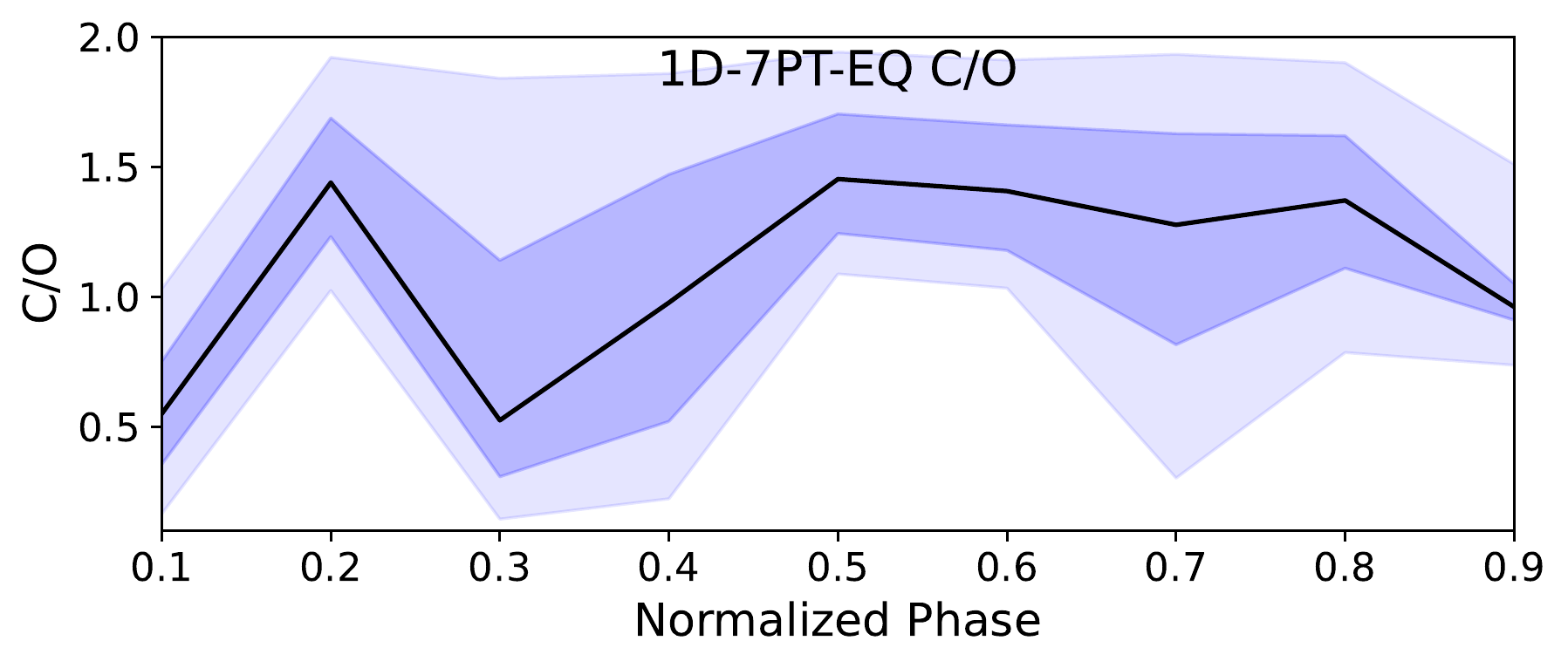}
    \caption{Recovered metallicity (top) and C/O (bottom) from the 1D-7PT-EQ retrievals. The 1$\sigma$ and 3$\sigma$ confidence intervals are shown by the shaded regions.}
    \label{fig:met_co}
\end{figure}

The retrieved parameters controlling the chemistry in the EQ runs are metallicity and C/O (see Figure \ref{fig:met_co}). When considering \emph{HST} and \emph{Spitzer} data, degeneracies exist between those two parameters, which is reflected in the retrieved posteriors for those parameters. Indeed, HST is mainly sensitive to water, which can be controlled by both metallicity and C/O, while the available two Spitzer bands are sensitive to CH$_4$, CO, and CO$_2$. The degeneracies in the Spitzer bands are exacerbated by the lack of baseline and redundancy for carbon-bearing species in HST. When HST and Spitzer are combined, studies have shown how instrument systematics can bias the abundance estimates for those species \citep{yip_lc, Yip_2021_w96}. In the EQ runs, the metallicity is consistent with solar, while super-solar C/O ratio seems to often be favored, but given the uncertainties, it remains hard to interpret those results.  

\begin{figure}
    \includegraphics[width = 0.5\textwidth]{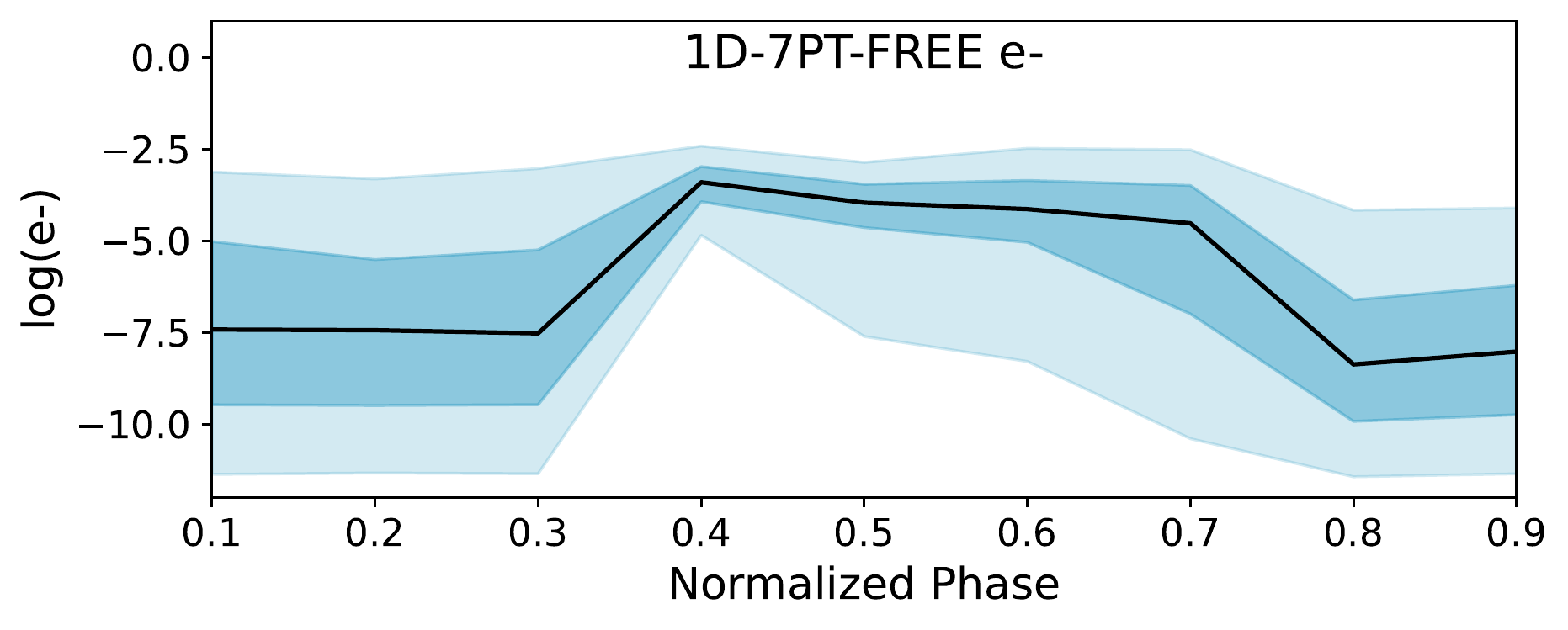}
    \includegraphics[width = 0.5\textwidth]{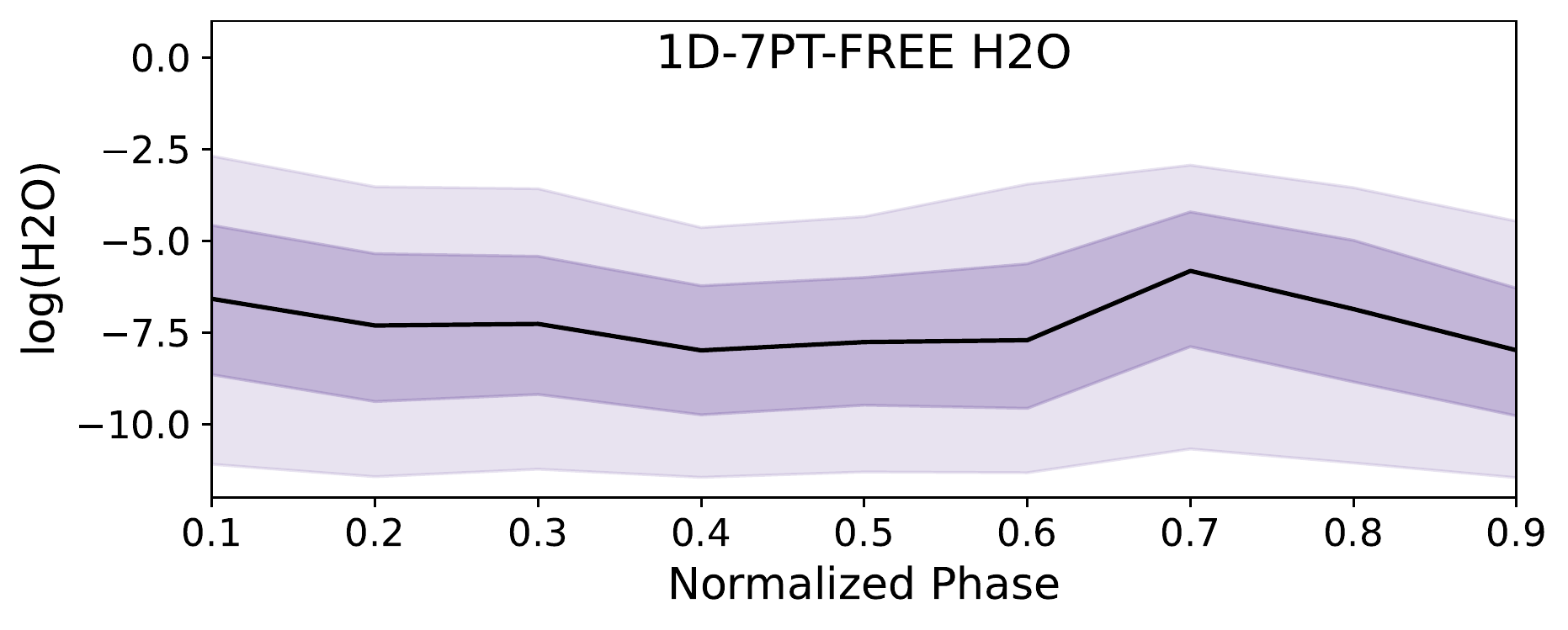}
    \caption{Recovered e$^-$ (Top) and H$_2$O (Bottom) abundances in the 1D-7PT-FREE retrievals. These results highlight the detection of H$^-$ opacity, marking the dissociation of H$_2$ and H$_2$O on the day-side.}
    \label{fig:1d_free_e-}
\end{figure}

To interpret the information content of the WASP-103\,b spectra, FREE retrievals are also explored. As shown in Figure \ref{fig:tp_map_free}, the retrieved thermal structure in the FREE case is similar to the EQ case. For the chemistry, since each species is recovered individually, it is possible to extract unbiased information regarding their location and abundances. In particular, Figure \ref{fig:1d_free_e-}, demonstrate the detection of e$^-$ opacity and an upper limit of H$_2$O, which is a strong confirmation that dissociation processes are indeed important for this atmosphere. The retrieved abundances for the other species are available in  Appendix 3. These show that only poor constraints can be inferred from individual phases on the carbon-bearing species, which is most likely due to the uncertainties on \emph{Spitzer} data and explains the large uncertainties on the C/O ratio and metallicity parameters in the EQ runs. Phase 0.5 exhibits some additional emission at 4.5$\mu$m, as noted by \cite{Kreidberg_w103}. In the equilibrium runs, this is better handled by an increase in C/O ratio via CO emission, while the free retrievals prefer a larger abundance of CO$_2$. From a statistical point of view, it is not possible to clearly distinguish between the two species, due to their identical contribution in this wavelength channel. For TiO, upper limits of about 10$^{-6}$ are obtained for the day-side. For the 1D runs, the Bayesian Evidence is in general similar between the different assumptions tested in this paper (see Appendix 4) but for most phases, a simple blackbody fit is not preferred. The retrievals around phase 0.5, which contains more information due to the higher emission of the planet, display significant evidence in favor of non-blackbody-like spectra ($\Delta log(E) > 5$).  

\subsection{1.5D unified retrieval}

\begin{figure}
    \includegraphics[width = 0.5\textwidth]{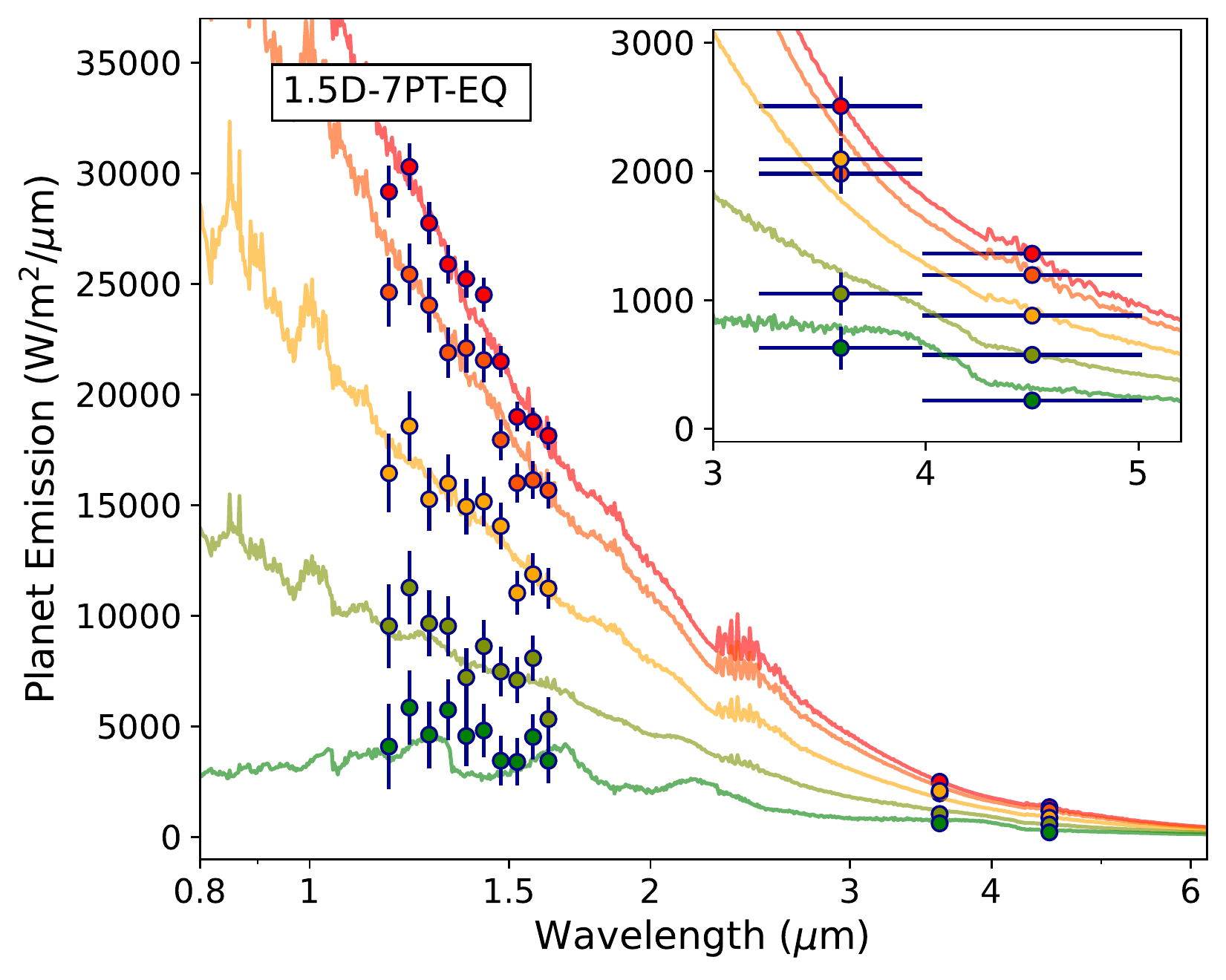}
    \includegraphics[width = 0.5\textwidth]{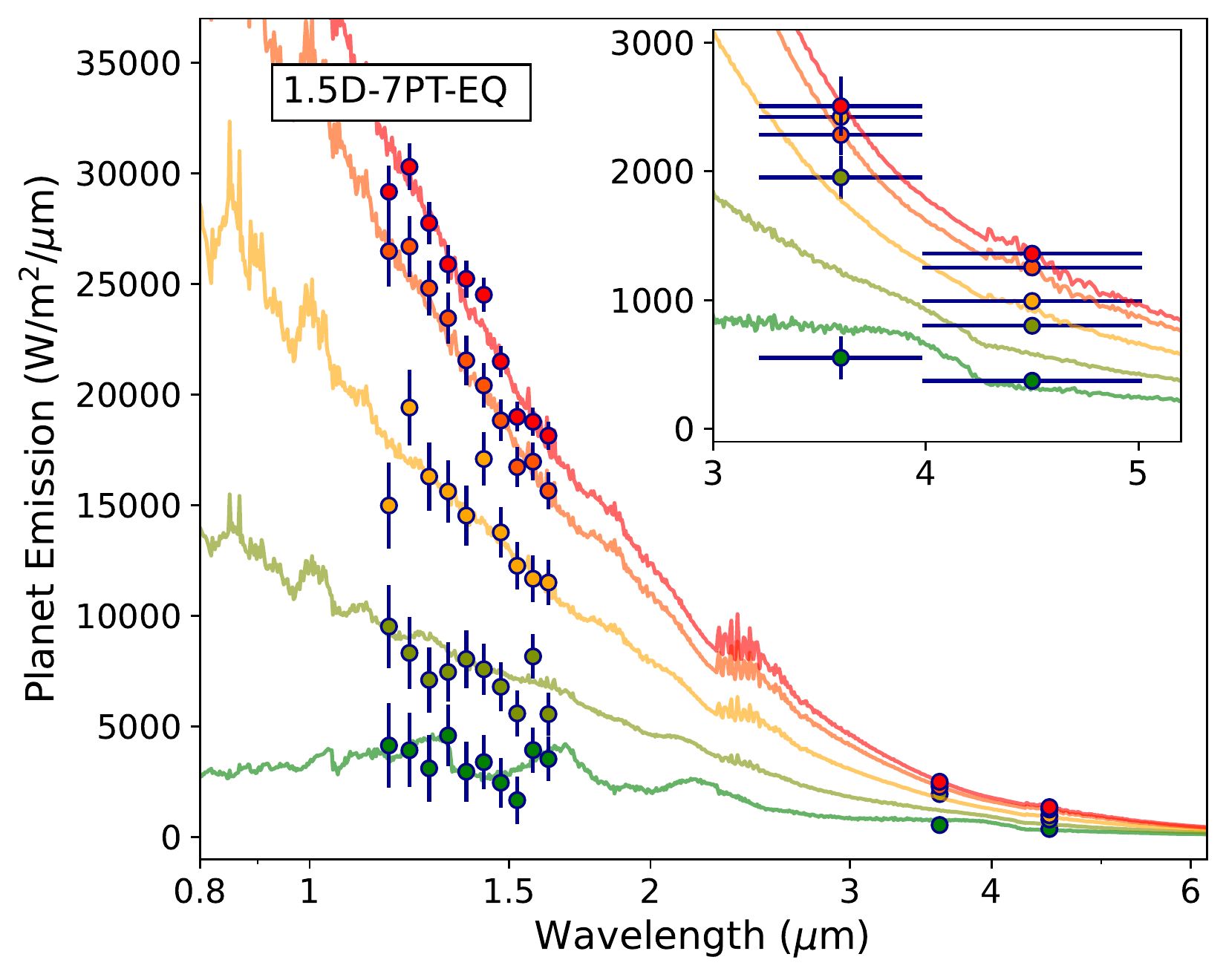}
    \caption{Observed and best fit spectra obtained in the 1.5D-7PT-EQ scenario. Top: phases from 0.1 to 0.5, from green to red; Bottom: phases from 0.5 to 0.9, from red to green.}
    \label{fig:1.5d_7pt_eq_spec}
\end{figure}

In the 1.5D retrieval, much better constraints on the properties of WASP-103\,b are expected as the information contained in each phase adds up to be fitted with a single representation of the planet's atmosphere. In this case, the atmosphere contains three homogeneous regions of different properties. In the first retrieval, the model assumes equilibrium chemistry, with metallicity and C/O ratio considered constant thorough the atmosphere. The observed and best-fit spectra for this retrieval, assuming the hot-spot size of 70 degrees, are shown in Figure \ref{fig:1.5d_7pt_eq_spec}. As seen in this figure, the retrieved model well follows the observations. However, as compared to the best-fit spectra obtained using the individual 1D approach, the model has more difficulties to fit the \emph{Spitzer} data. This is due to the observed \emph{Spitzer} phase-curve having some very scattered datapoints, in particular, the 0.7 and 0.8 phase observations at 3.6$\mu$m, which have much higher emission than the model predictions. In the individual approaches, since the models are independent, this could have impacted the detection of carbon-bearing species. In the unified 1.5D retrieval, since all the phases are fit together, the retrieval should not be driven by individual scattered datapoints, and one can expect more reliable results regarding carbon-bearing species.

\begin{figure}
    \centering
    \includegraphics[width = 0.32\textwidth]{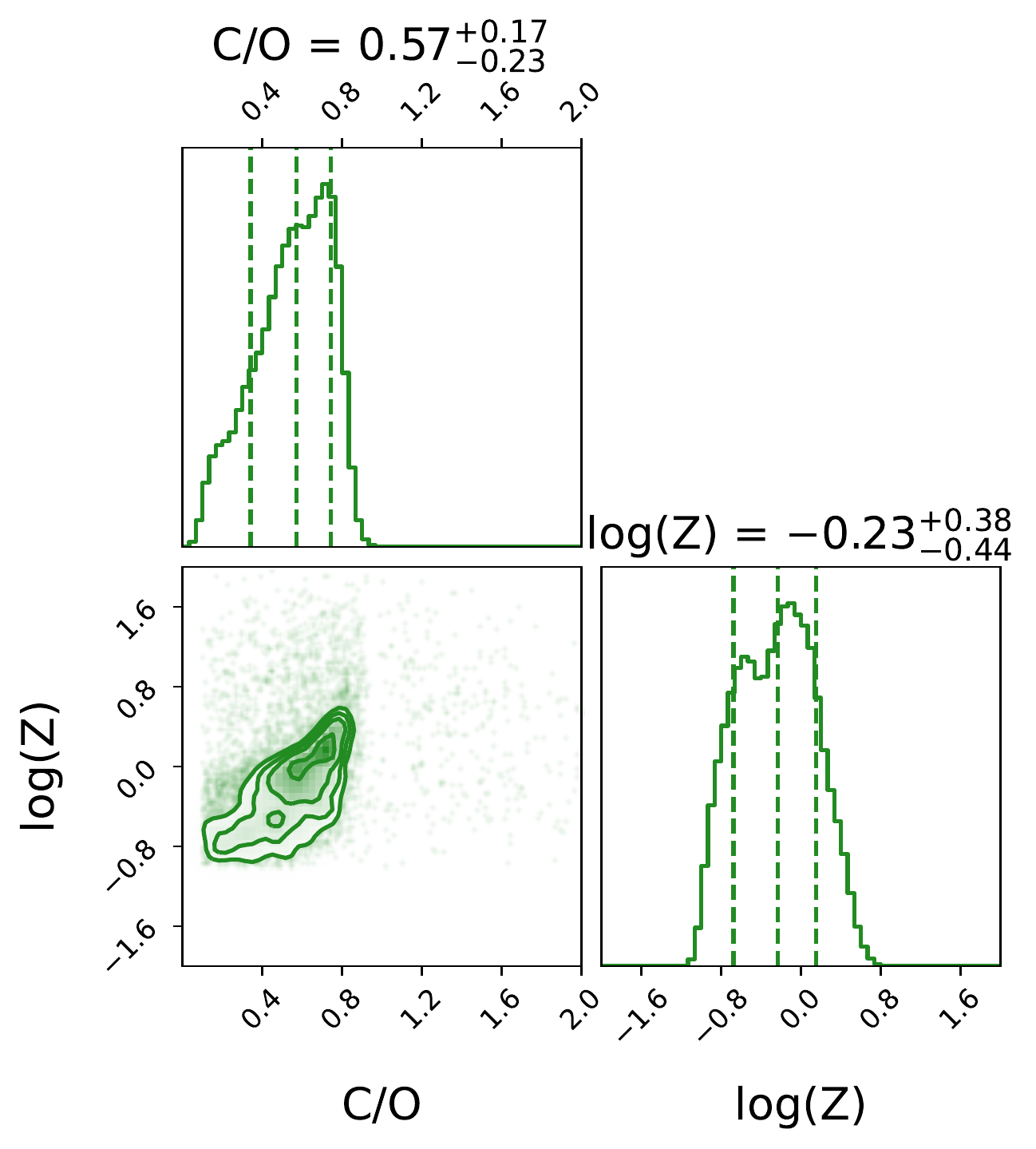}
    \includegraphics[width = 0.33\textwidth]{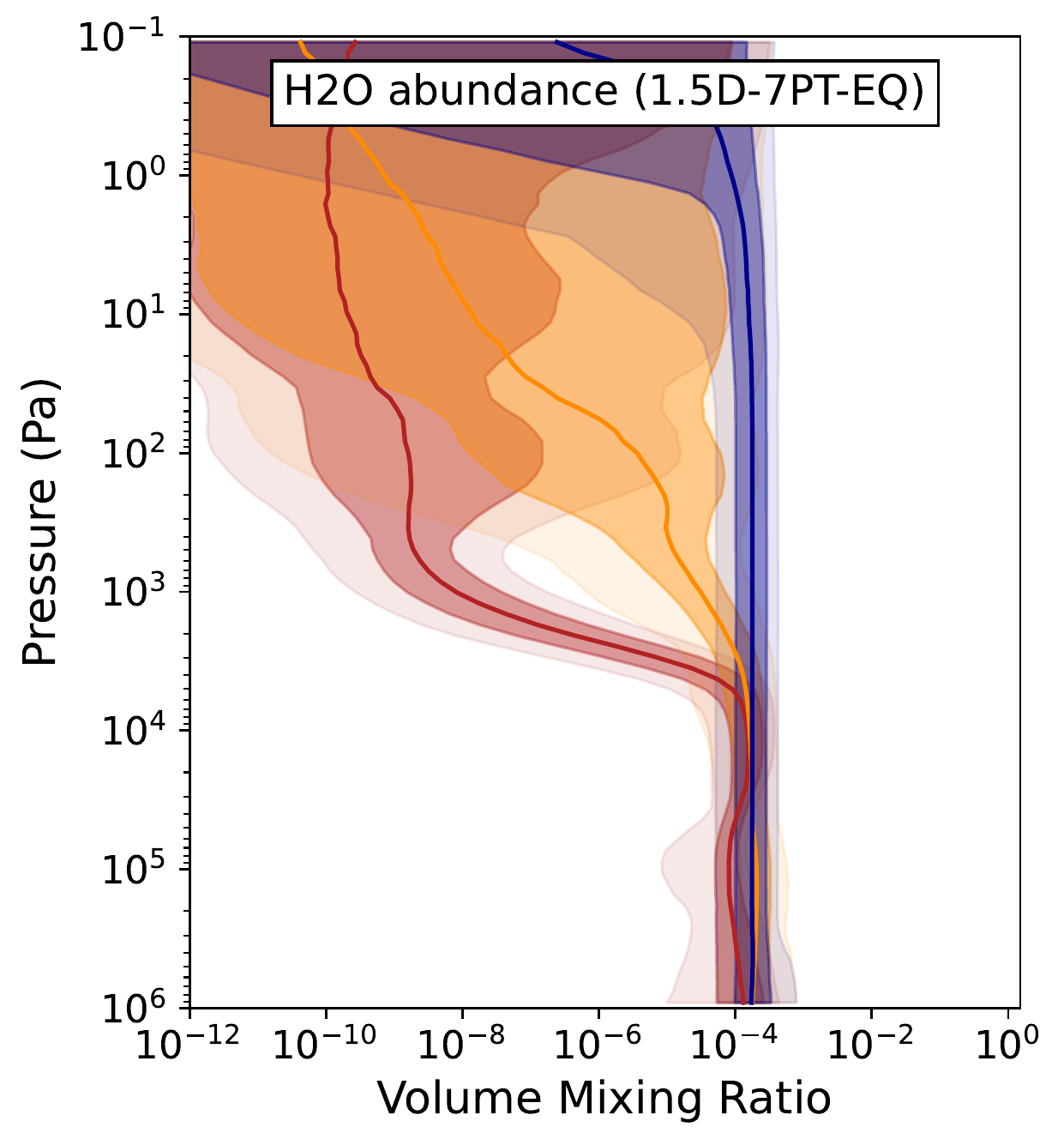}
    \includegraphics[width = 0.32\textwidth]{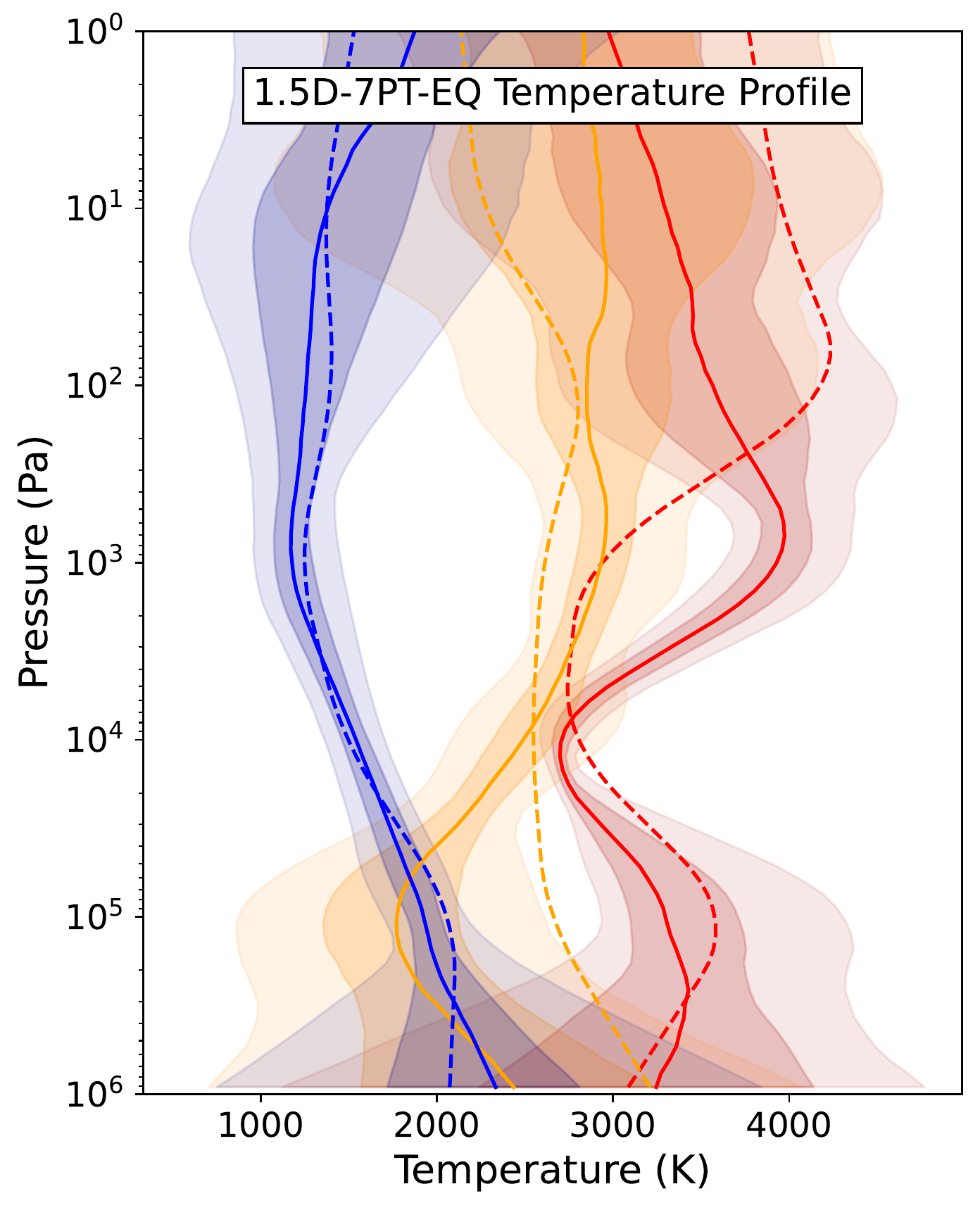}
    \caption{Posterior distribution (Top), water abundance profile (Middle) and temperature-pressure profile (Bottom) obtained in the 1.5D-7PT-EQ unified retrieval. Red: hot-spot; Orange: day-side; blue: night side. The best-fit temperature profiles of the 1D-7PT-EQ runs for phases 0.5 (red), 0.7 (orange) and 0.9 (blue) are also provided in dashed-lines for comparison. }
    \label{fig:1.5d_7pt_eq_post}
\end{figure}

The 1.5D-7PT-EQ retrieval indicates that the data is consistent with a solar metallicity (log Z = -0.23$^{+0.38}_{-0.44}$) and a solar C/O (C/O = 0.57$^{+0.17}_{-0.23}$) atmosphere. This is seen in Figure \ref{fig:1.5d_7pt_eq_post}, which also presents the volume mixing ratio of H$_2$O and the Temperature-Pressure structure in each region. Large uncertainties remain in the recovered chemistry since the spectra do not show strong molecular features. The day-side is consistent with a thermal inversion at 10$^4$ Pa with temperatures reaching 4000K, as also demonstrated in the 1D runs. On the day-side and hot-spot regions, this thermal inversion is likely enhanced by dissociation of H$_2$, which leads to strong absorbing properties in the visible due to H$^-$. This behavior is clearly seen in the chemical profiles derived from the equilibrium chemistry model, which are available for each molecule in Appendix 5. 

In the second retrieval, the information content of the data is explored further by assuming free chemistry with constant with altitude profiles. Since this assumption involves a much larger parameter space, with a free parameter for each species in each region, the Temperature-Pressure profile is fixed to the mean from the previous 1.5D-7PT-EQ run. The retrieved posteriors for the chemical species are reported in Appendix 6. As expected the H$^-$ opacity, parametrized by the abundance of e$^-$ is retrieved on the hot-spot,  providing direct indications that thermal dissociation is important in this atmosphere. The depletion of H$_2$O can also be characterized, with H$_2$O being detected in all regions, with abundances as low as 10$^-5$. On the hot-spot, CO is detected from the additional emission observed near the normalized phase 0.5 at 4.5$\mu$m, which was already highlighted in \cite{Kreidberg_w103}. TiO and VO are not detected in this atmosphere, but FeH, which is more thermally stable is found on the hot-spot and the day-side, with abundances that are consistent with the equilibrium retrievals. At phases near 0.25 and 0.75, an offset is observed between the 3.6$\mu$m and 4.5$\mu$m channels, which is fit using emission of CH$_4$ on the day-side. This is surprising as at those temperatures, the dominant carbon-bearing species are CO and CO$_2$, but in the model, only CH$_4$ can provide the additional emission at 3.6$\mu$m. For C-bearing species, since the retrieved abundances rely on photometric points, using a more constrained approach with a single parameter for all the species as is done in the EQ retrieval might be preferable.

A comparison of the retrieved best-fit spectra and thermal structures for the 1D-7PT-EQ and the 1.5D-7PT-EQ is available in an animated figure (digital version of the manuscript) in Appendix 7.

\section{Discussion}

\subsection{Comparison with previous results}

Overall, the results presented here are consistent with the main findings of \cite{Kreidberg_w103}. However, as compared to the previous analysis and by using a new methodology, this study extracts a richer and more statistically robust picture for the atmosphere of WASP-103\,b. This work improves on the determination of the metallicity and the C/O ratio for this planet, which is found consistent with solar, stressing the importance of using a wide range of techniques (1D, 1.5D, free chemistry, equilibrium chemistry) to ensure robust estimates. In the JWST era, such a cautious approach, employing various assumptions and cross-checking their validity, will be crucial as the higher signal-to-noise and broader wavelength coverage offered by this instrument will be sensitive to many new processes that are likely to bias a given analysis technique. To best extract information content from phase-curve data, a unified retrieval approach, sensitive to the 3D nature of these observations and encompassing all the spectra at once, is demonstrated to be more accurate. Using such technique, the precise thermal structure of WASP-103\,b, as well as the origins for the signals of various molecules (H$_2$O, FeH, CO and CH$_4$) and for dissociation processes, can be extracted and mapped to the different regions of this atmosphere. 

\subsection{Use of individual and unified retrievals}

From a modeling point of view, phase-curve data are notoriously more difficult to analyze than the more traditional transit and eclipse observations due to their sensitivity to more complex processes. The individual 1D retrieval technique is a simple adaptation of an eclipse retrieval that does not require significant compute time \citep{Changeat_2020_pc2}. However, since each observation is retrieved individually, assuming no correlation with the other observations, contamination from emission at other phases is almost guaranteed \citep{Feng_2016, Taylor_2020}. Such effects have motivated the development of tools to correct for the contamination \citep{Cubillos_2021}. In general, 1D retrievals on phase-curve data should allow to obtain first constraints and guide more complex techniques, such as the 1.5D retrievals presented here. In the 1.5D, or any other unified technique, the emission at each phase is computed according to the weight of the different regions, which means that contamination does not occur. The information available in each spectrum and contributing to each phase is accounted for, so better constraints on each region can also be obtained. One disadvantage, however, is that the dimension of the retrieval can increase rapidly and require larger computing resources. Both 1D and 1.5D approaches can be used to check the consistency of the results, as it is done in this study. 

To go a step further, one could use the constraints obtained from 1D and 1.5D retrievals to generate Global Circulation Models (GCM) of the planet, checking both the interpretation of the spectra and the level of understanding regarding the physico-chemical processes included in the GCM.

\subsection{Transit: stellar activity, instrument combination and atmospheric variabilities}

\begin{figure}
    \centering
    \includegraphics[width = 0.49\textwidth]{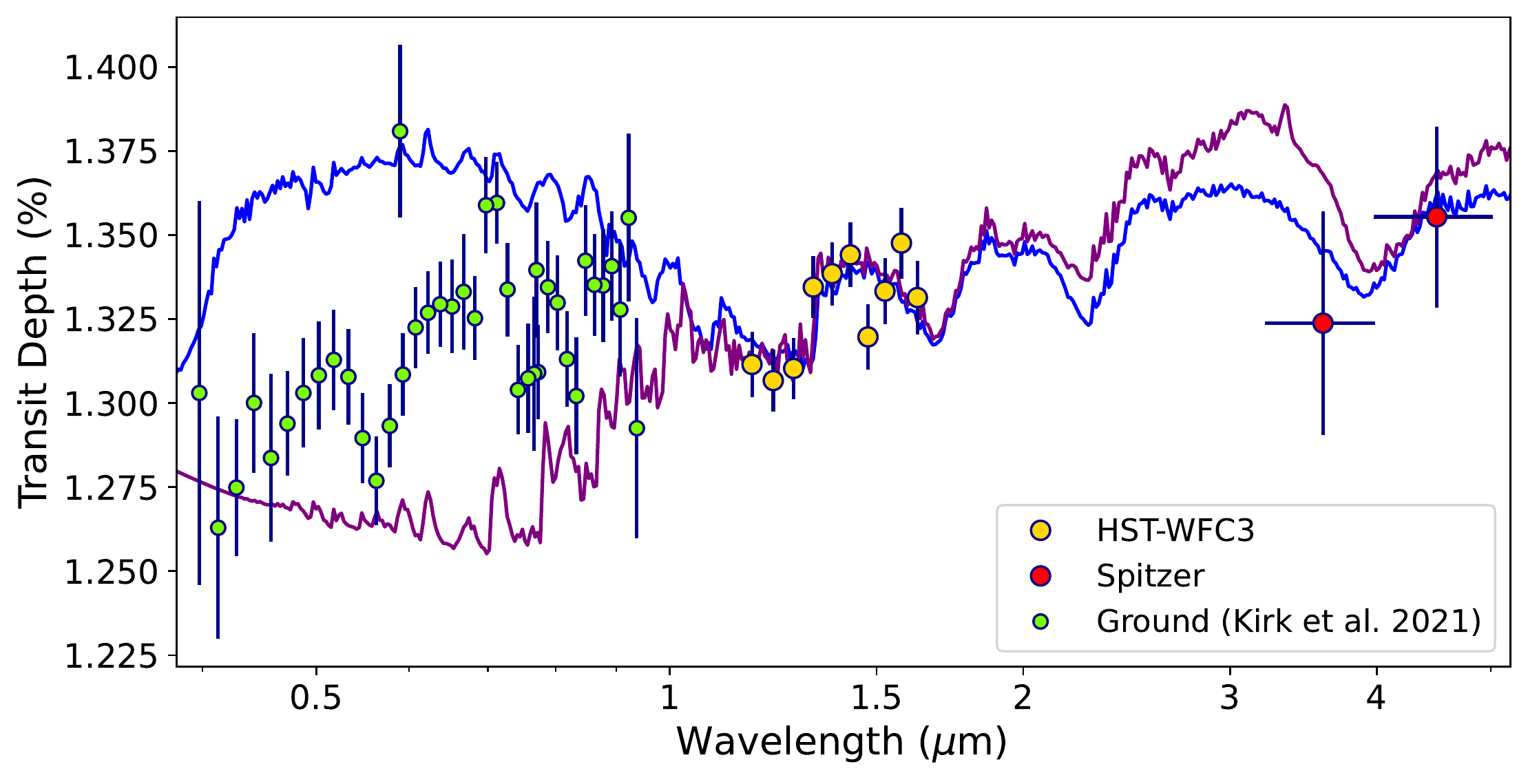}
    \caption{Observed and best-fit spectra for the primary transit (phase 0.0) in the 1.5D models. Blue: 7PT-EQ; Purple: 7PT-FREE. The data from  \cite{Kirk_2021_w103}, which is not included in the retrievals, is shown for comparison. }\label{fig:15d_transits}
\end{figure}

The transit observations from \emph{HST} and \emph{Spitzer} were also included in the 1D and 1.5D retrievals. The transit geometry is more sensitive to the planetary radius than phase observations,  while also providing complementary information on the terminator region. Including the transit observations in the 1.5D retrievals leads to very tight constraints on the radius (1.55$\pm$0.01). The \emph{HST} transit observation is consistent with the absorption of water vapor at 1.4$\mu$m, as seen in Figure \ref{fig:15d_transits} and as found in 1D retrievals of the transit. 

For this planet, additional ground-based observations exist \citep{Kirk_2021_w103}, but they were not included in the retrievals presented here due to potential incompatibilities. When running a 1D retrieval on the complete transit spectra, results were found to be non-physical with abundances of TiO reaching values about 10$^{-3}$. This result highlights the difficulties of combining observations from different instruments \citep{yip_lc, Yip_2021_w96}. \cite{Kirk_2021_w103} included light-source effects in their retrieval analyses, finding that such effects could explain the downward slope in the combined \emph{ACCESS}, \emph{LRG-BEASTS}, \emph{Gemini/GMOS} and \emph{VLT/FORS2} observations. This result is however surprising as WASP-103 is an F8 dwarf \citep{gillon_wasp103} for which no stellar activity was explicitly reported. For this star, photometric modulations of 5 mmag were measured \citep{Kreidberg_w103}. While the potential contamination from stellar activity might, in any case, be less important in the infrared observations considered here, the  combination of different instruments could introduce biases, especially as observations from different epochs are combined (\emph{HST}-WFC3, \emph{Spitzer} 3.6$\mu$m and \emph{Spitzer} 4.5$\mu$m). 
Similarly, \cite{Cho_2021} used spectral methods to demonstrate the complex behavior of 3-dimensional atmospheres. In their study of atmospheric dynamics, storms and modons develop from small-scale instabilities, which would lead to significant temporal variabilities. Such effects would in principle render the analysis of data taken at different times extremely difficult.    

\section{Conclusion}

The retrievals performed in this study (1D and 1.5D) provide similar, consistent results. They indicate that the WASP-103\,b phase-curve data is best fit with a thermal inversion on the day-side. The inversion is associated with thermal dissociation processes, which can be tracked via the e$^-$ and H$_2$O abundances. The presence of carbon-bearing species cannot clearly be confirmed from the 1D retrievals, but the 1.5D retrievals, which by design are more efficient at extracting information content from entire phase-curve data, are able to put constraints on the metallicity and the C/O ratio for this planet. The planet is consistent with solar values and the spectral features of H$_2$O, H$^-$, FeH, CO, and CH$_4$ are found in different regions of WASP-103\,b. H$_2$O for instance is extracted in all regions of the atmosphere, including the night-side. Overall, this work demonstrates the relevance of unified methods, such as the 1.5D phase-curve retrieval, and their complementarity with more traditional 1D models. Phase-curve observations have the potential to provide a better understanding of exoplanet atmospheres by breaking the 3D biases from single transits or eclipse observations. This understanding will be required to place reliable constraints on elemental abundances and therefore inform planetary formation and evolution models. 

\vspace{3mm}

\software{TauREx3 \citep{2019_al-refaie_taurex3, al-refaie_2021_taurex3.1}, GGChem \citep{Woitke_2018}, Astropy \citep{astropy}, h5py \citep{hdf5_collette}, Matplotlib \citep{Hunter_matplotlib}, Multinest \citep{Feroz_multinest,buchner_multinest}, Pandas \citep{mckinney_pandas}, Numpy \citep{oliphant_numpy}, SciPy \citep{scipy}, corner \citep{corner}}.

\vspace{3mm}
\textbf{Data:} This work is based upon observations with the NASA/ESA Hubble Space Telescope, obtained at the Space Telescope Science Institute (STScI) operated by AURA, Inc. The data used in this publication was reduced in \cite{Kreidberg_w103}.

\vspace{3mm}
\textbf{Acknowledgements:} This project has received funding from the European Research Council (ERC) under the European Union's Horizon 2020 research and innovation programme (grant agreement No 758892, ExoAI), from the Science and Technology Funding Council grants ST/S002634/1 and ST/T001836/1 and from the UK Space Agency grant ST/W00254X/1. The author thanks Giovanna Tinetti, Ingo P. Waldmann and Ahmed F. Al-Refaie, as well as the anonymous reviewer for their useful recommendations and discussions.

This work utilised the OzSTAR national facility at Swinburne University of Technology. The OzSTAR program receives funding in part from the Astronomy National Collaborative Research Infrastructure Strategy (NCRIS) allocation provided by the Australian Government. This work utilised resources provided by the Cambridge Service for Data Driven Discovery (CSD3) operated by the University of Cambridge Research Computing Service (www.csd3.cam.ac.uk), provided by Dell EMC and Intel using Tier-2 funding from the Engineering and Physical Sciences Research Council (capital grant EP/P020259/1), and DiRAC funding from the Science and Technology Facilities Council (www.dirac.ac.uk).

\bibliographystyle{aasjournal}
\bibliography{main}

\renewcommand{\floatpagefraction}{.99}%

\section*{Supplementary Materials}

\subsection*{Appendix 1: Observed and best-fit spectra obtained in the 1D-7PT-FREE and the 1D-7PT-EQ retrievals}

\begin{figure}[h]
    \centering
    \includegraphics[width = 0.49\textwidth]{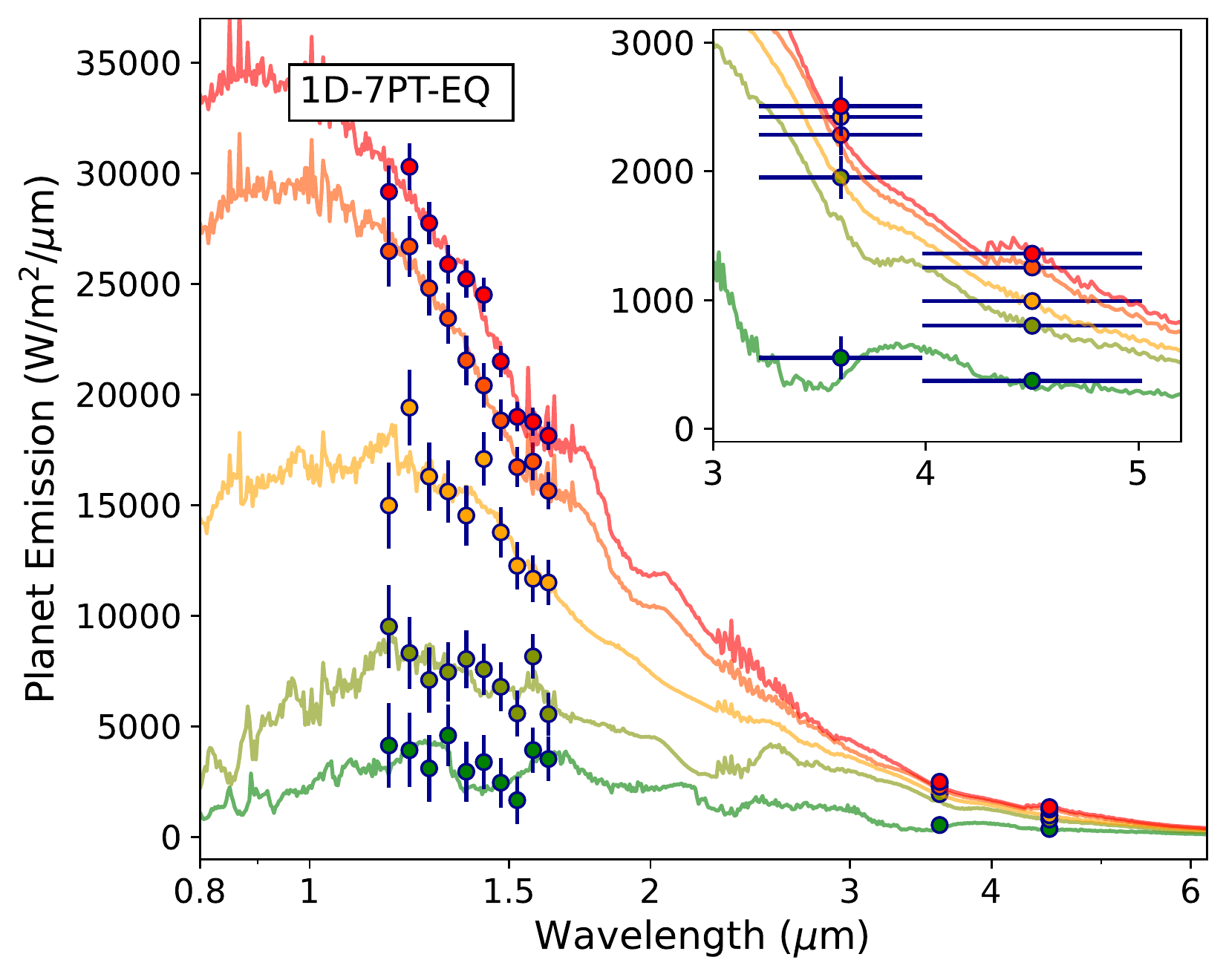}
    \includegraphics[width = 0.49\textwidth]{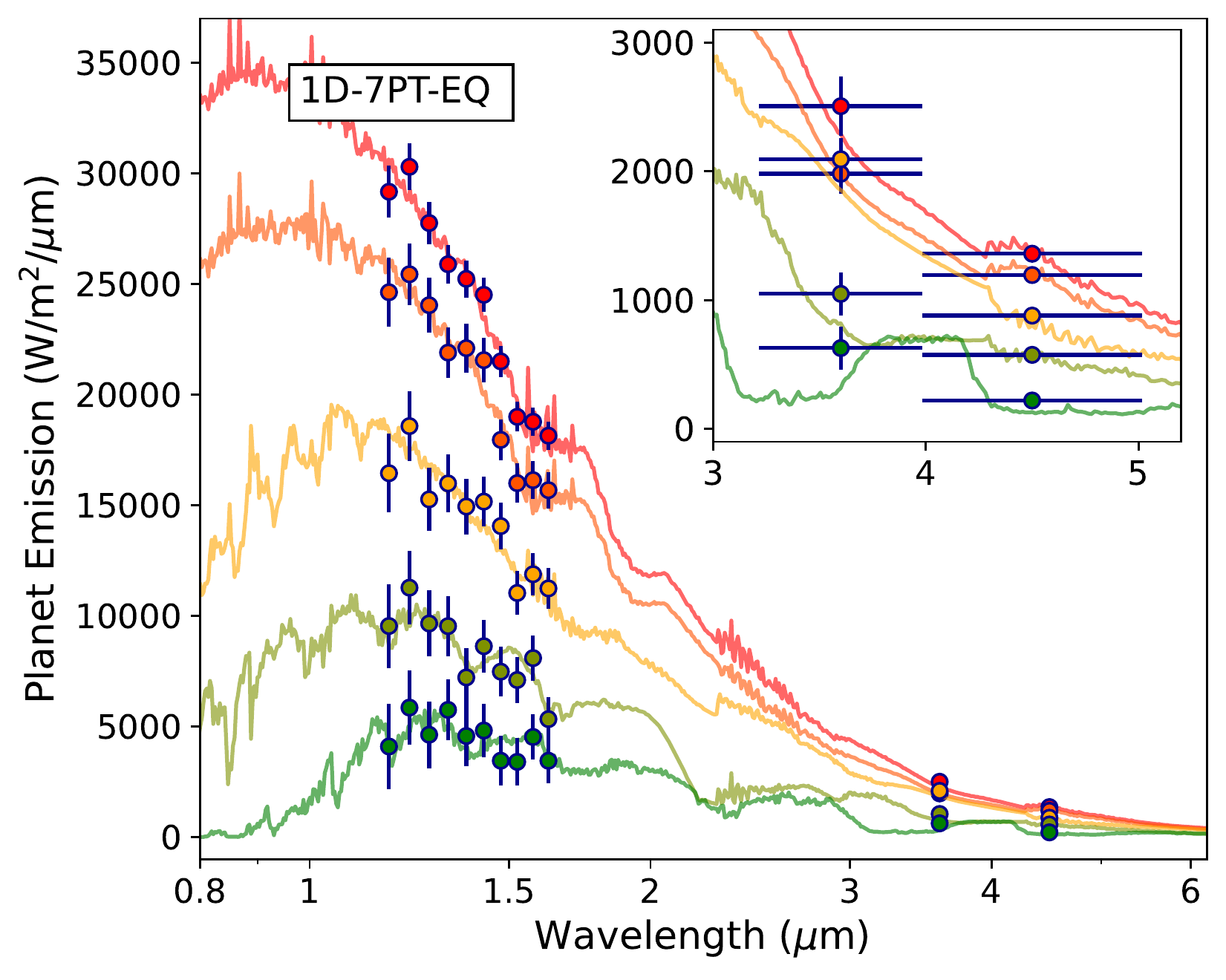}
    \includegraphics[width = 0.49\textwidth]{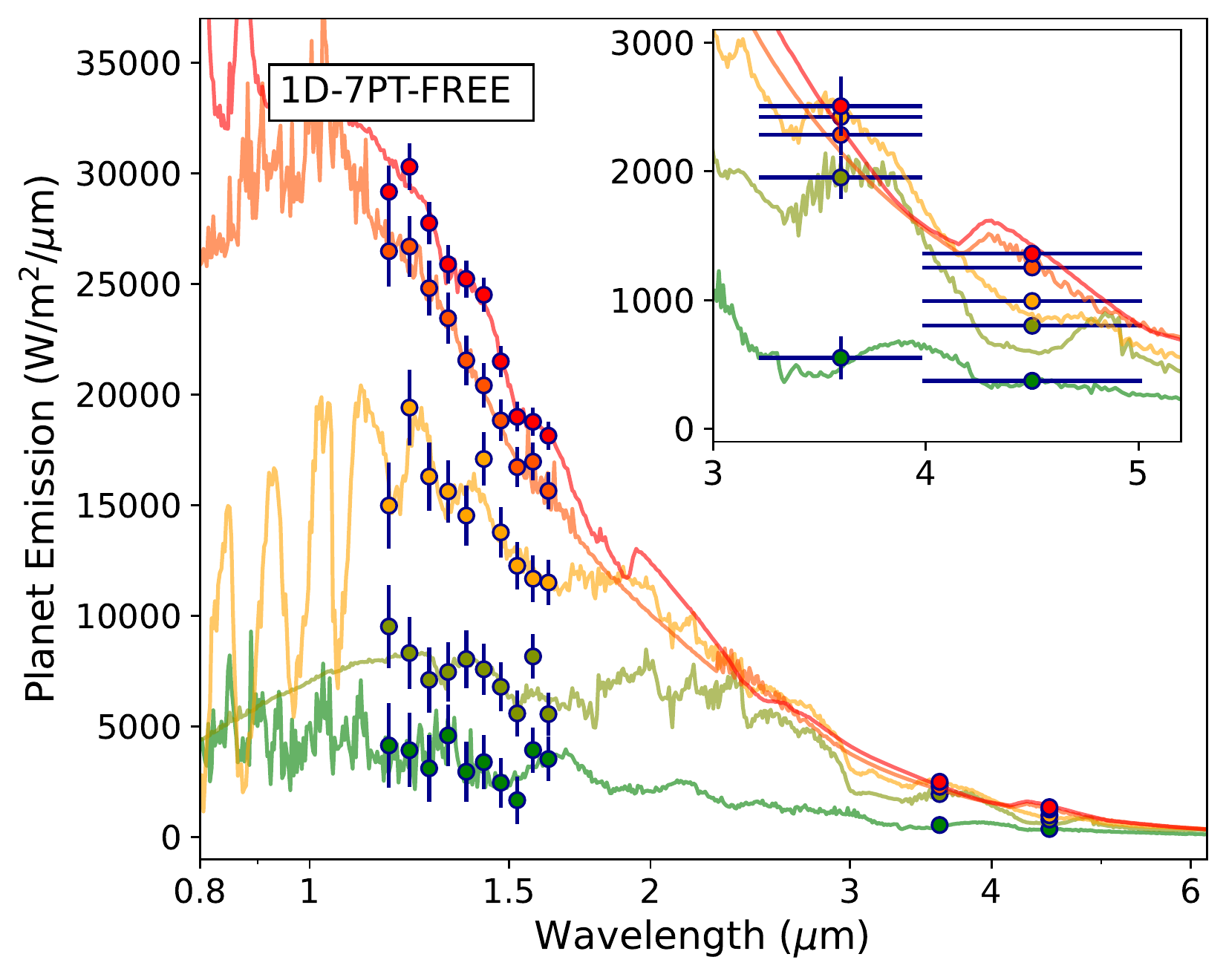}
    \includegraphics[width = 0.49\textwidth]{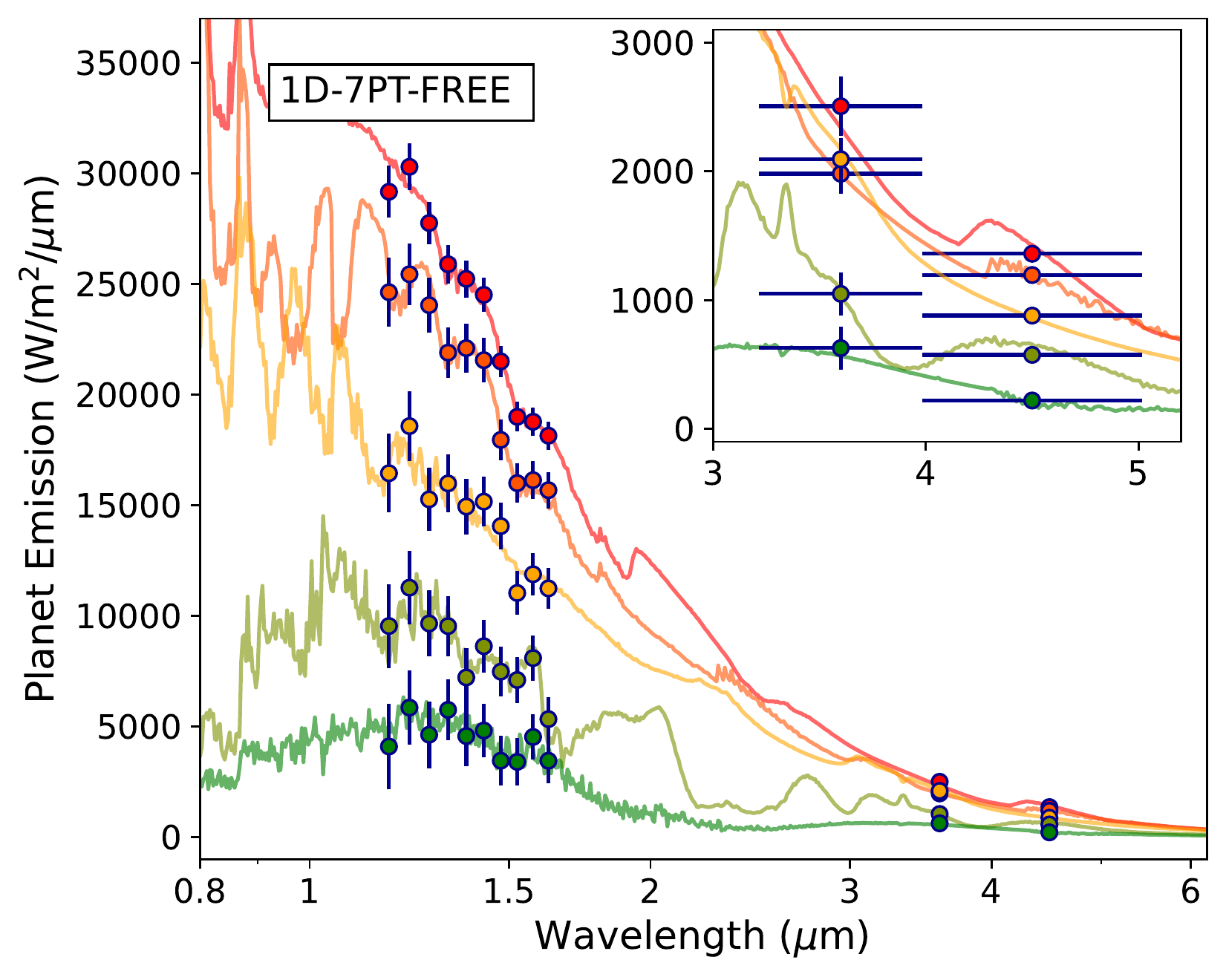}
    \caption{Observed and best fit spectra obtained in the 1D-7PT-EQ (Top) and the 1D-7PT-FREE retrievals. Left column: phases from 0.1 to 0.5, from green to red; Right column: phases from 0.5 to 0.9, from red to green.}\label{fig:1d_7pt_spectra}
\end{figure}

Figure \ref{fig:1d_7pt_spectra} displays the observed and best fit spectra of WASP-103\,b phase-curve from the 1D-7PT-EQ and the 1D-7PT-FREE models. 

\clearpage

\subsection*{Appendix 2: Complementary chemistry maps to the 1D-7PT-EQ retrievals}

\begin{figure}[h]
    \centering
    \includegraphics[width = 0.46\textwidth]{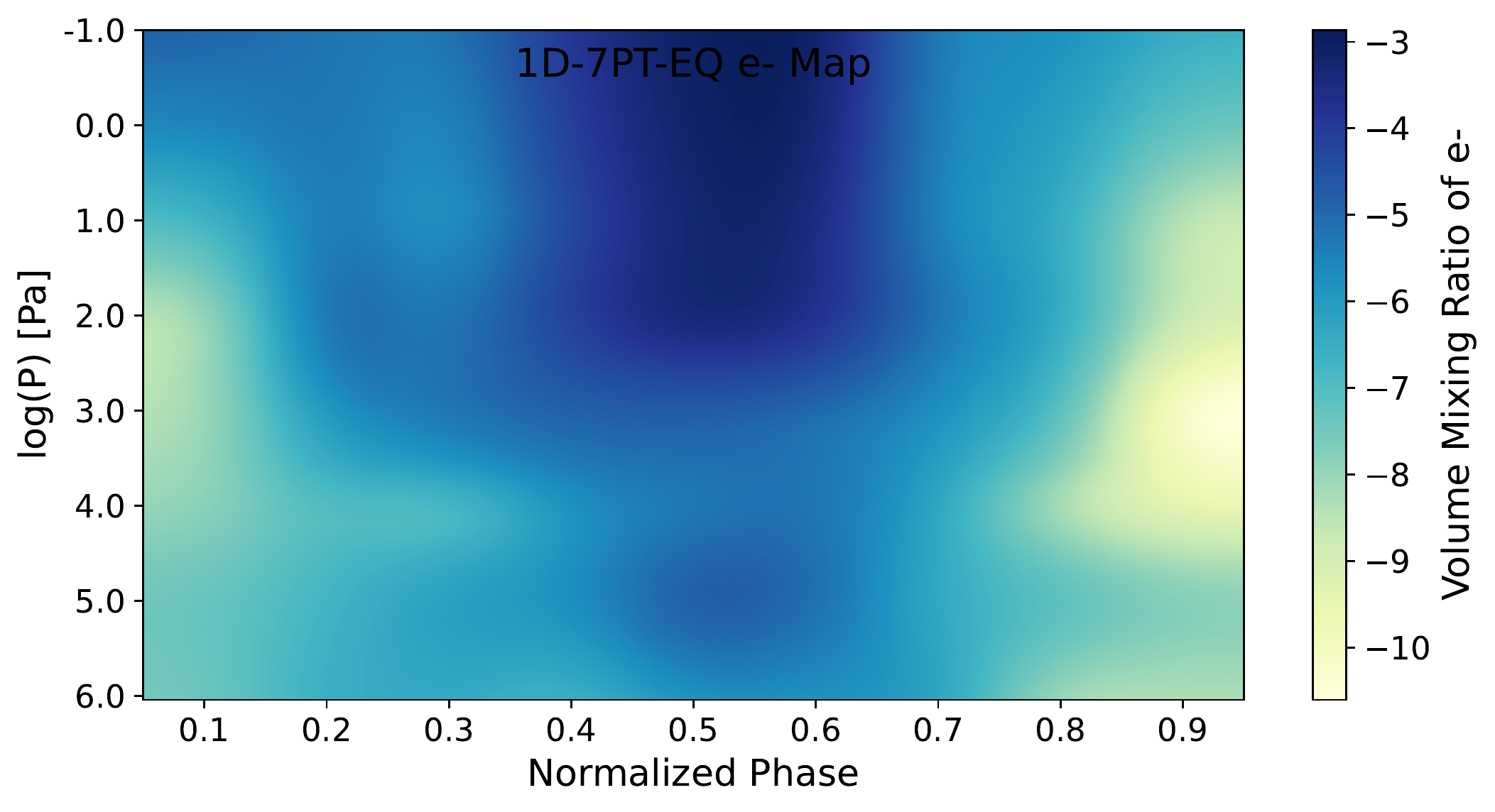}
    \includegraphics[width = 0.46\textwidth]{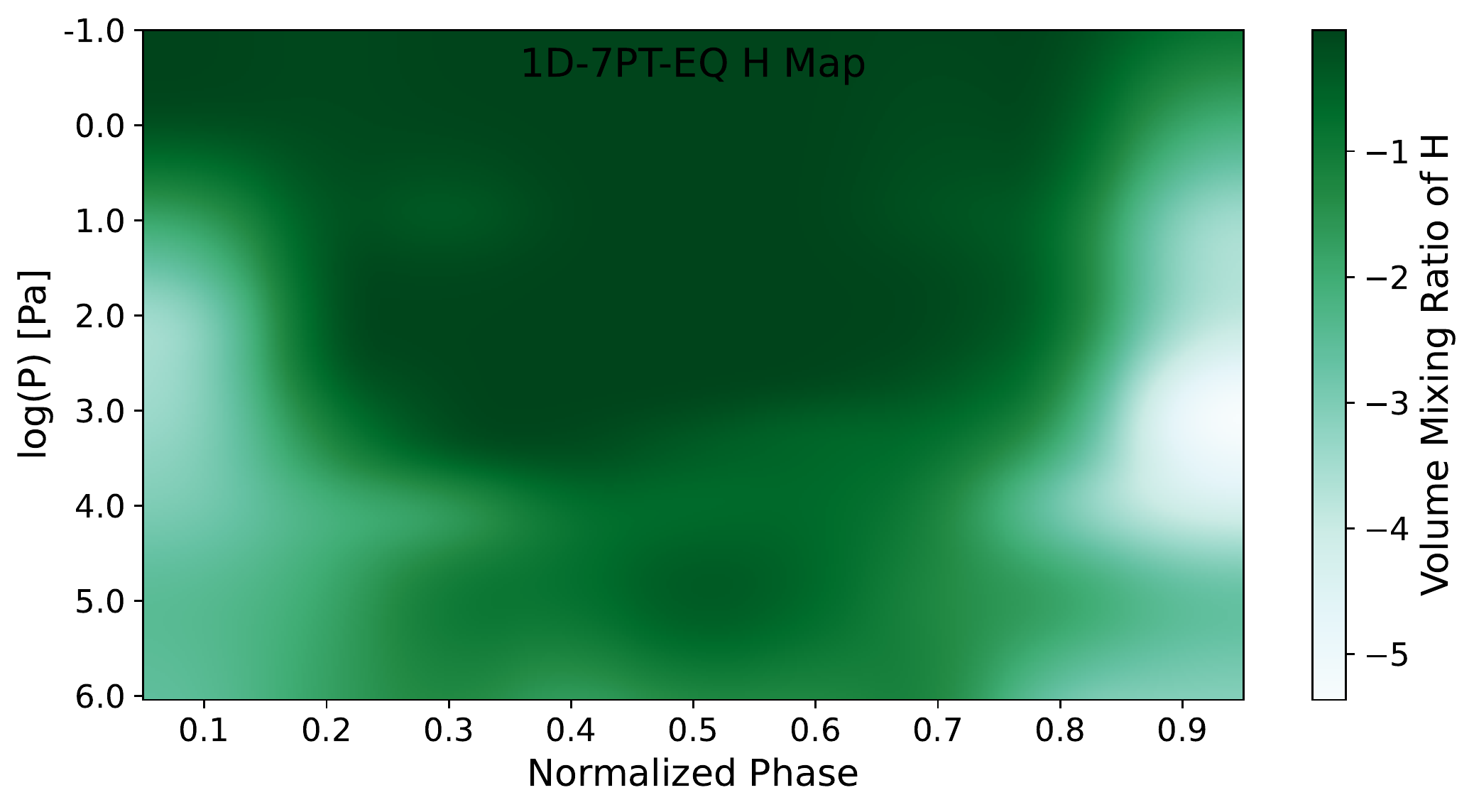}
    \includegraphics[width = 0.46\textwidth]{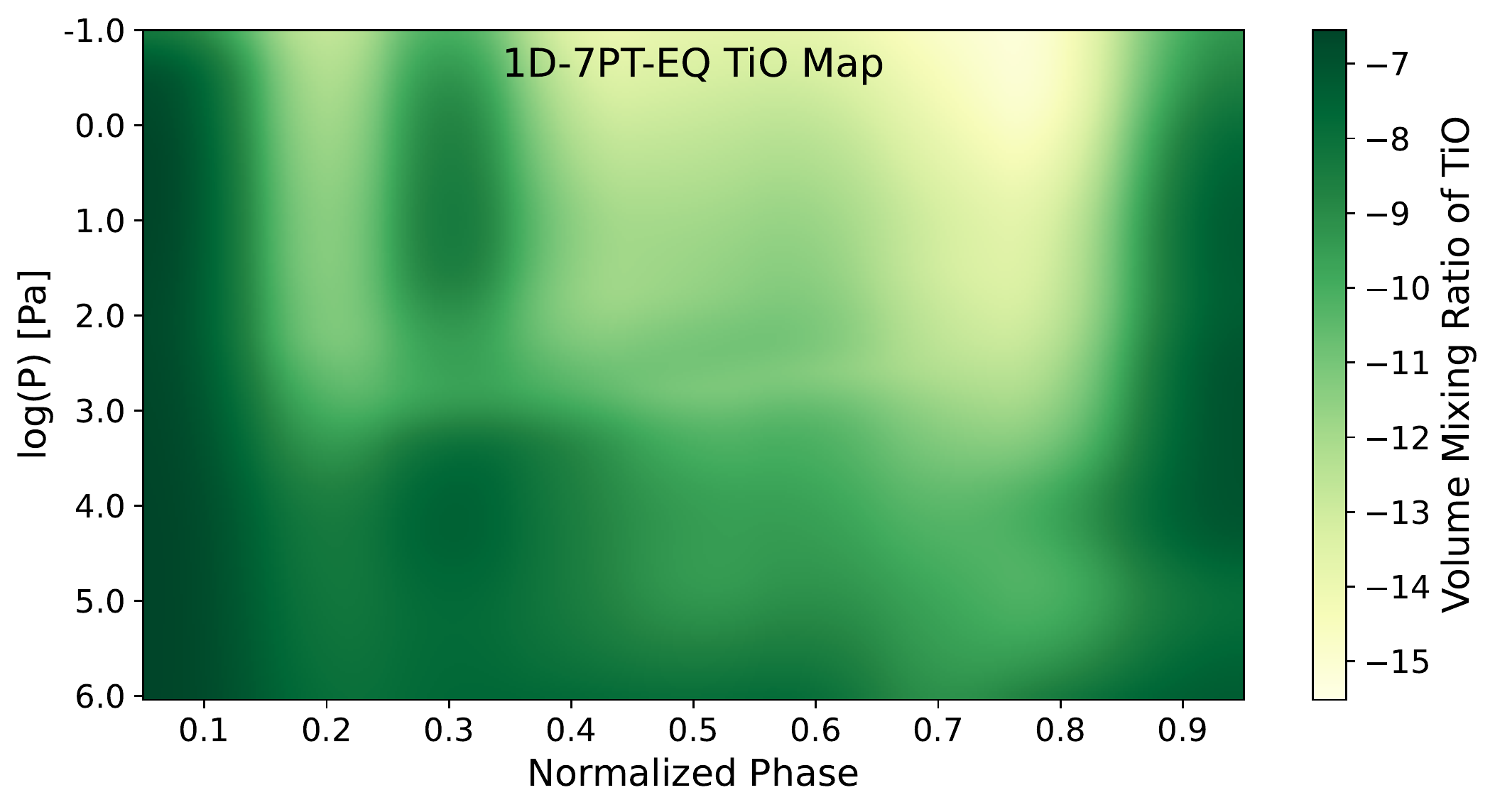}
    \includegraphics[width = 0.46\textwidth]{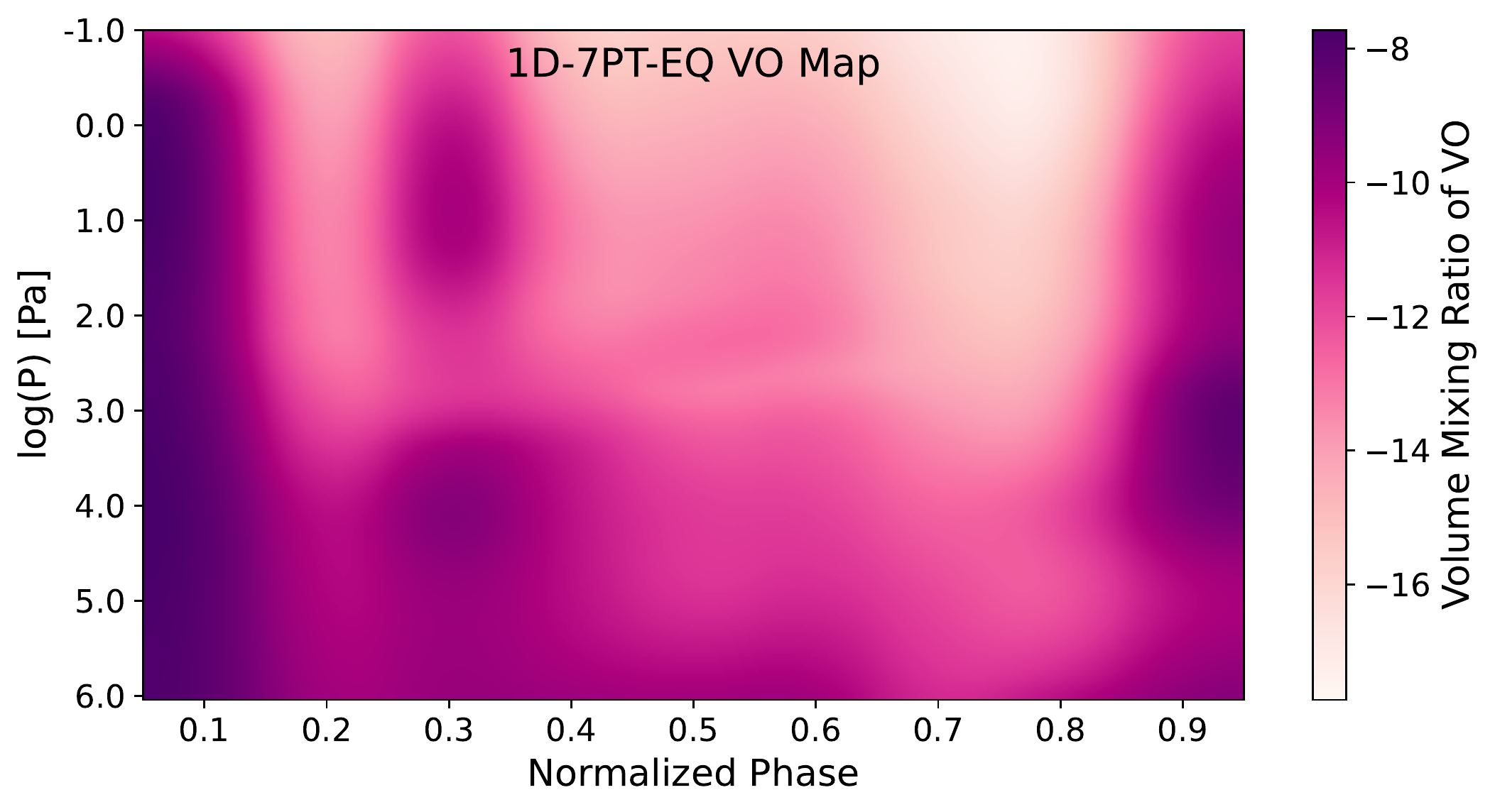}
    \includegraphics[width = 0.46\textwidth]{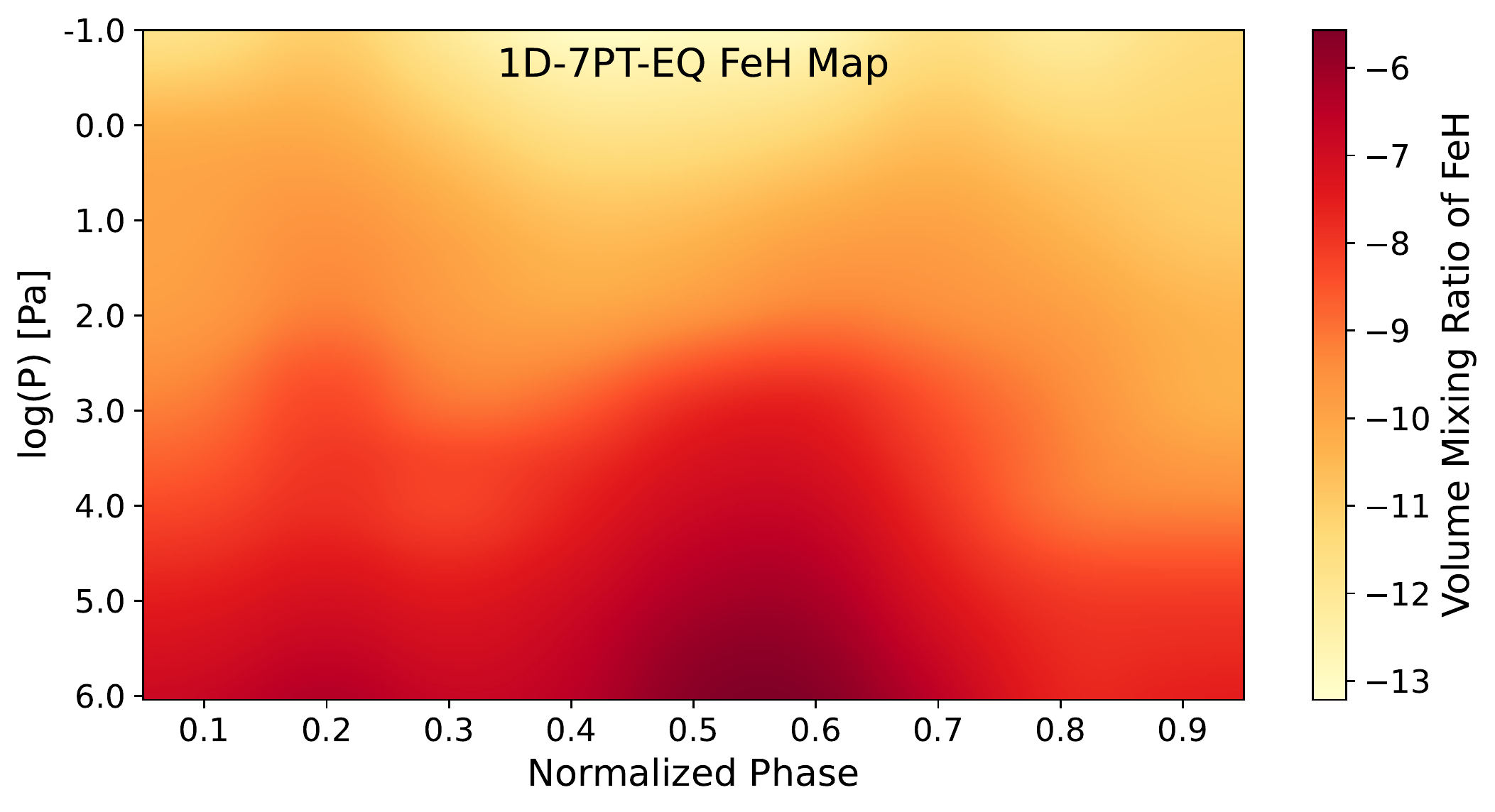}
    \includegraphics[width = 0.46\textwidth]{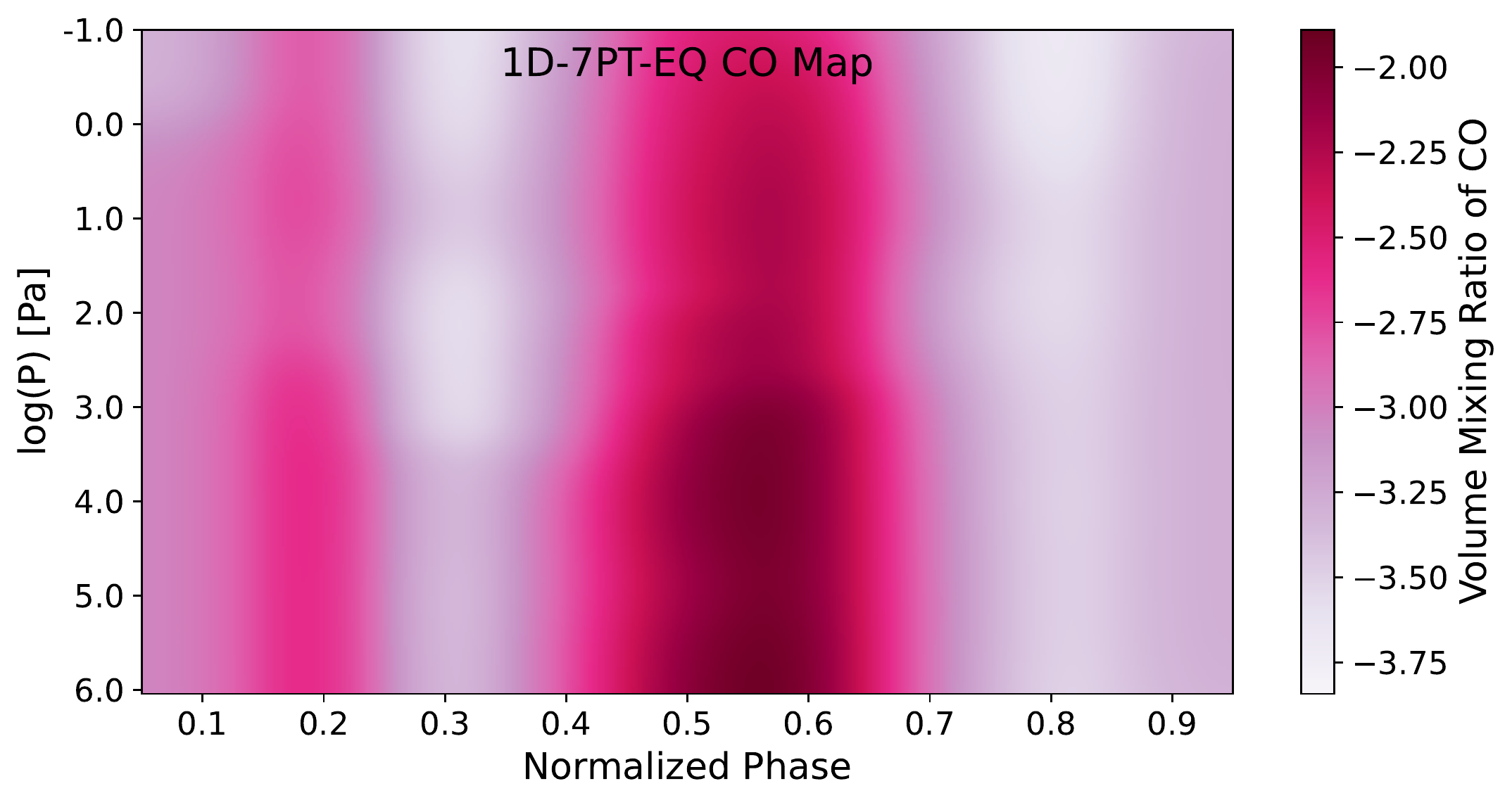}
    \includegraphics[width = 0.46\textwidth]{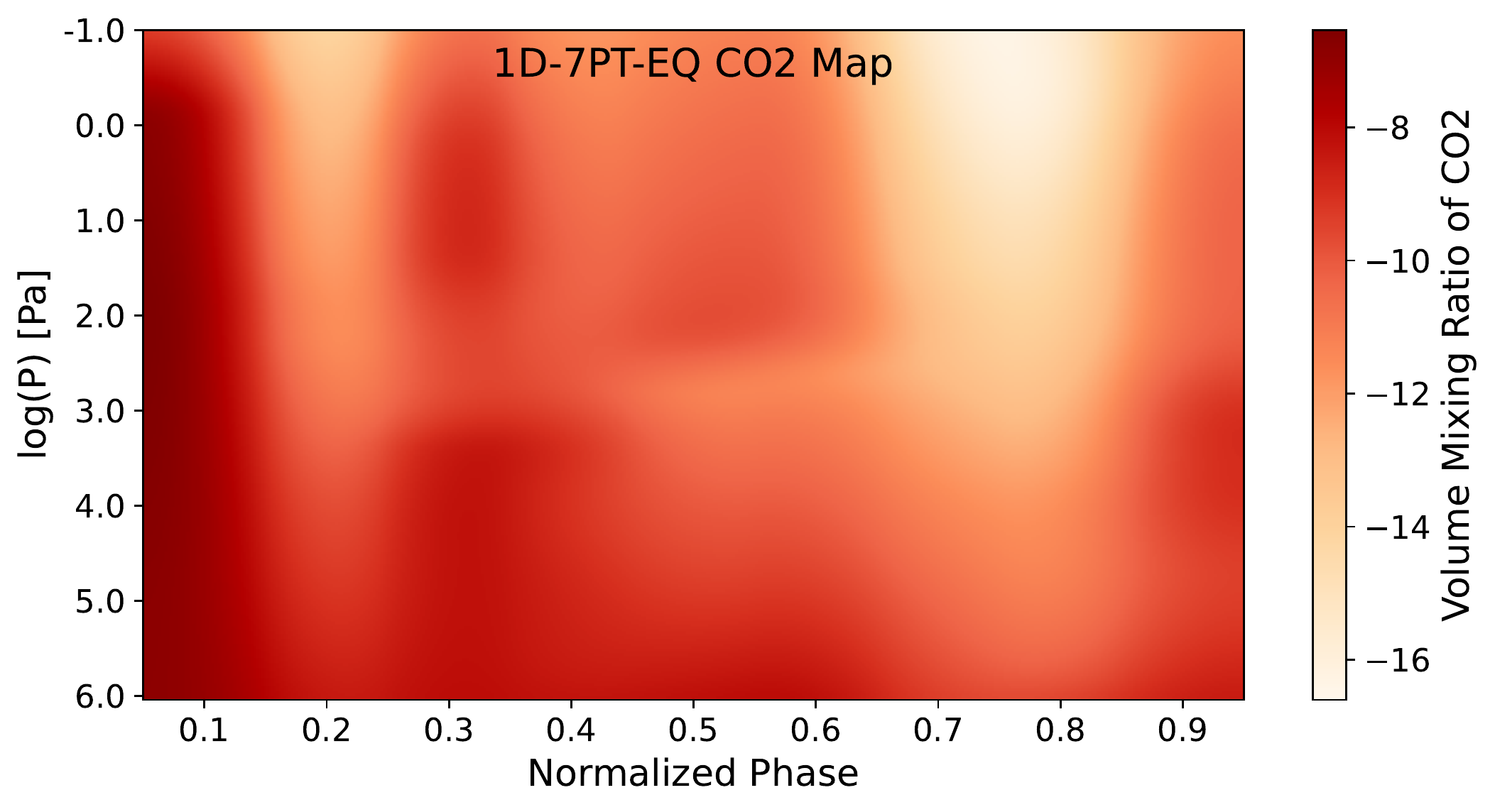}
    \includegraphics[width = 0.46\textwidth]{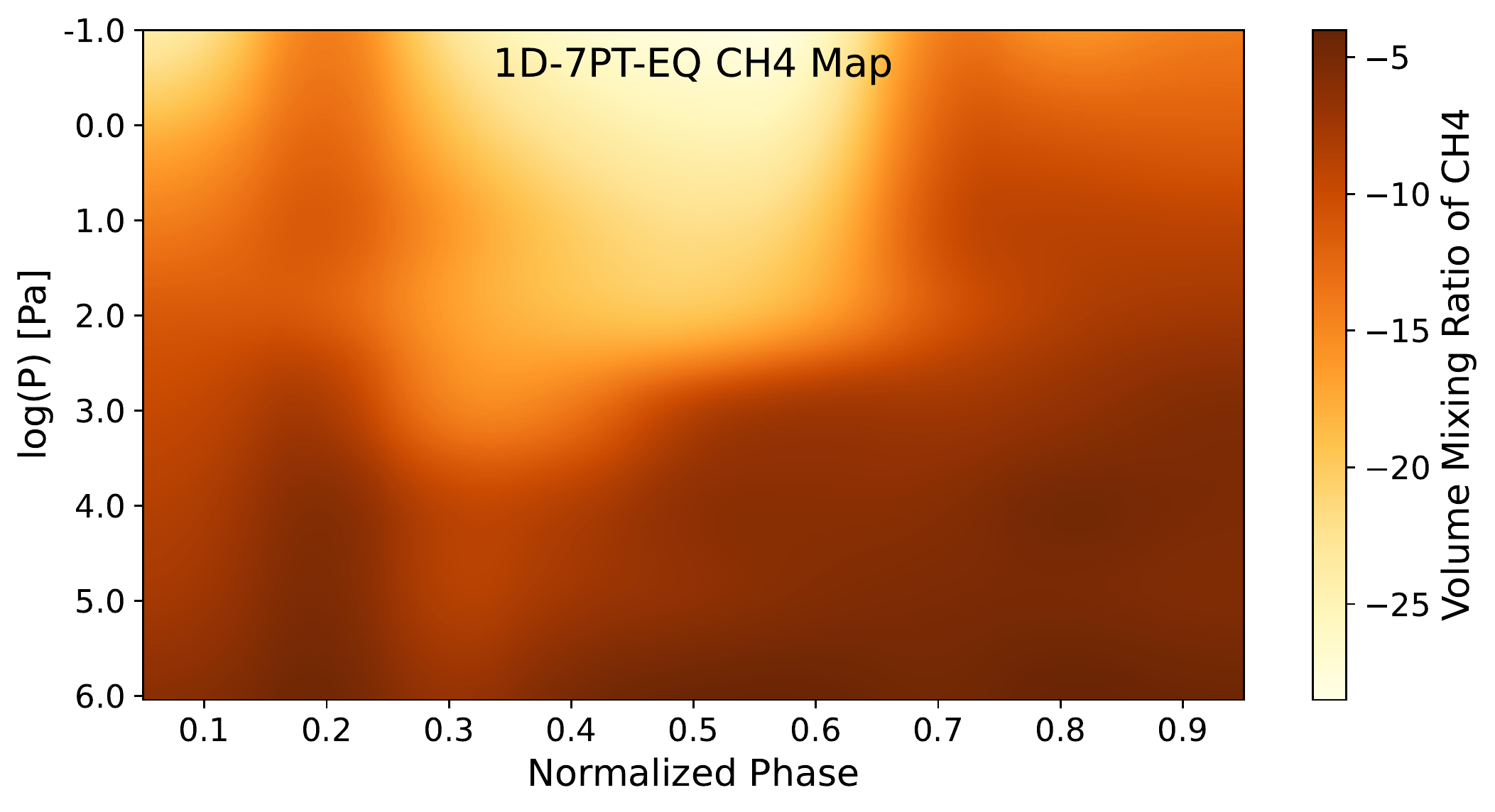}
    
    \caption{Recovered mean abundances from the 1D-7PT-EQ retrievals. First column in descending order: e$^-$, TiO, FeH and CO$_2$; Second column: H, VO, CO and CH$_4$.}\label{fig:e-_h2_map_eq}
\end{figure}

Figure \ref{fig:e-_h2_map_eq} shows the recovered abundance distributions as a function of phase and altitude for the main molecules in the 1D-7PT-EQ retrievals.

\clearpage

\subsection*{Appendix 3: Complementary retrieved abundances to the 1D-7PT-FREE retrievals}

\begin{figure}[h]
    \centering
    \includegraphics[width = 0.46\textwidth]{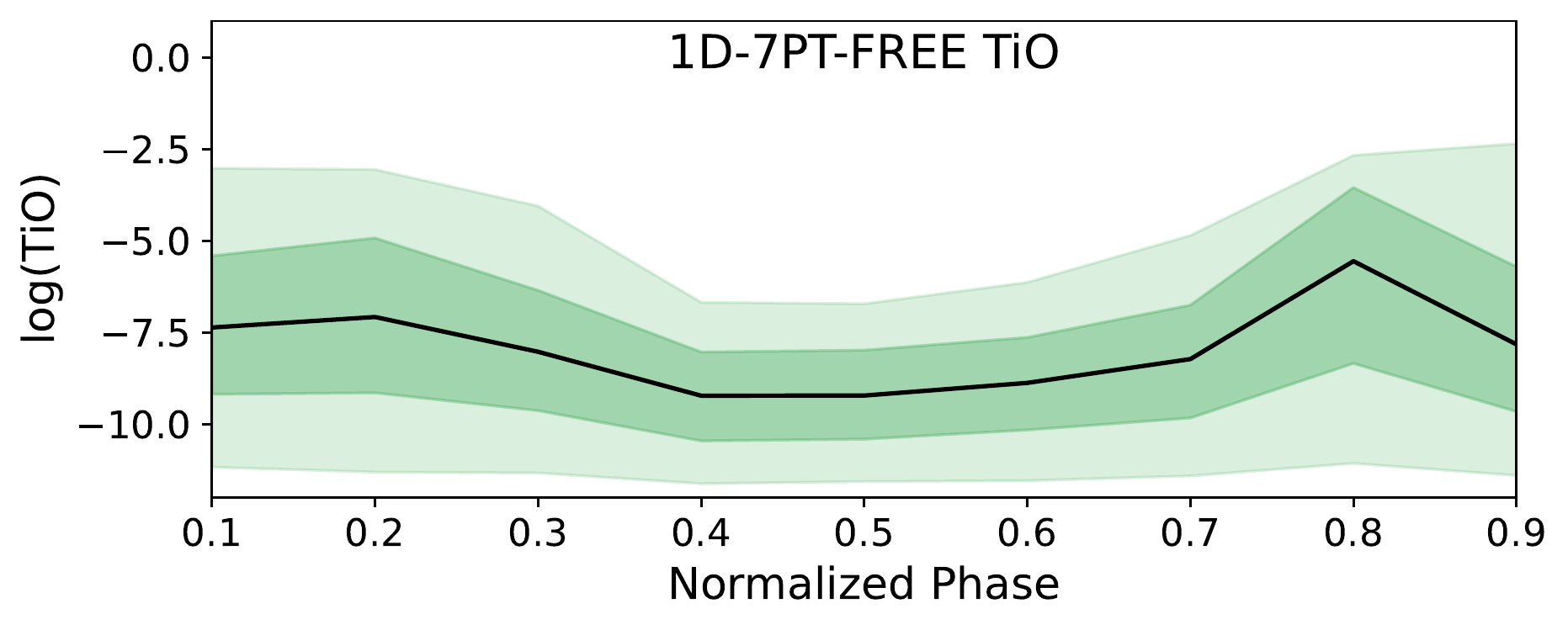}
    \includegraphics[width = 0.46\textwidth]{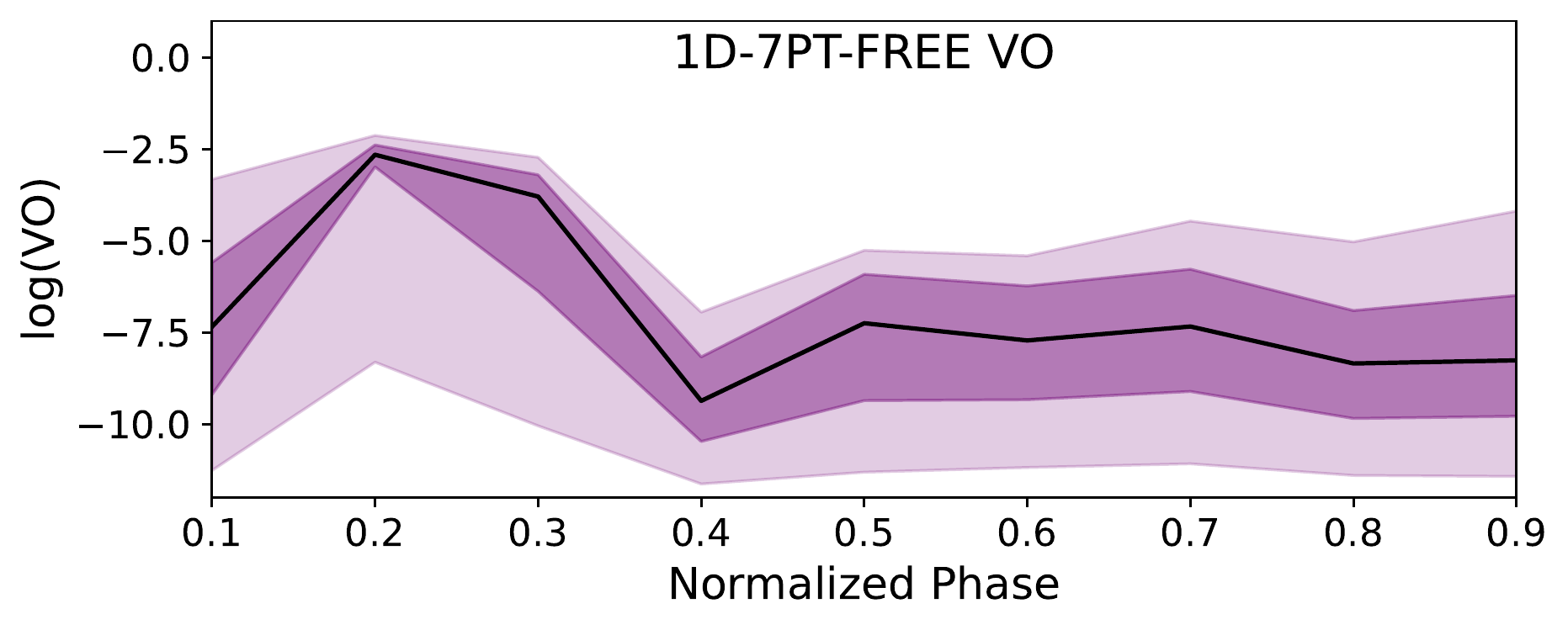}
    \includegraphics[width = 0.46\textwidth]{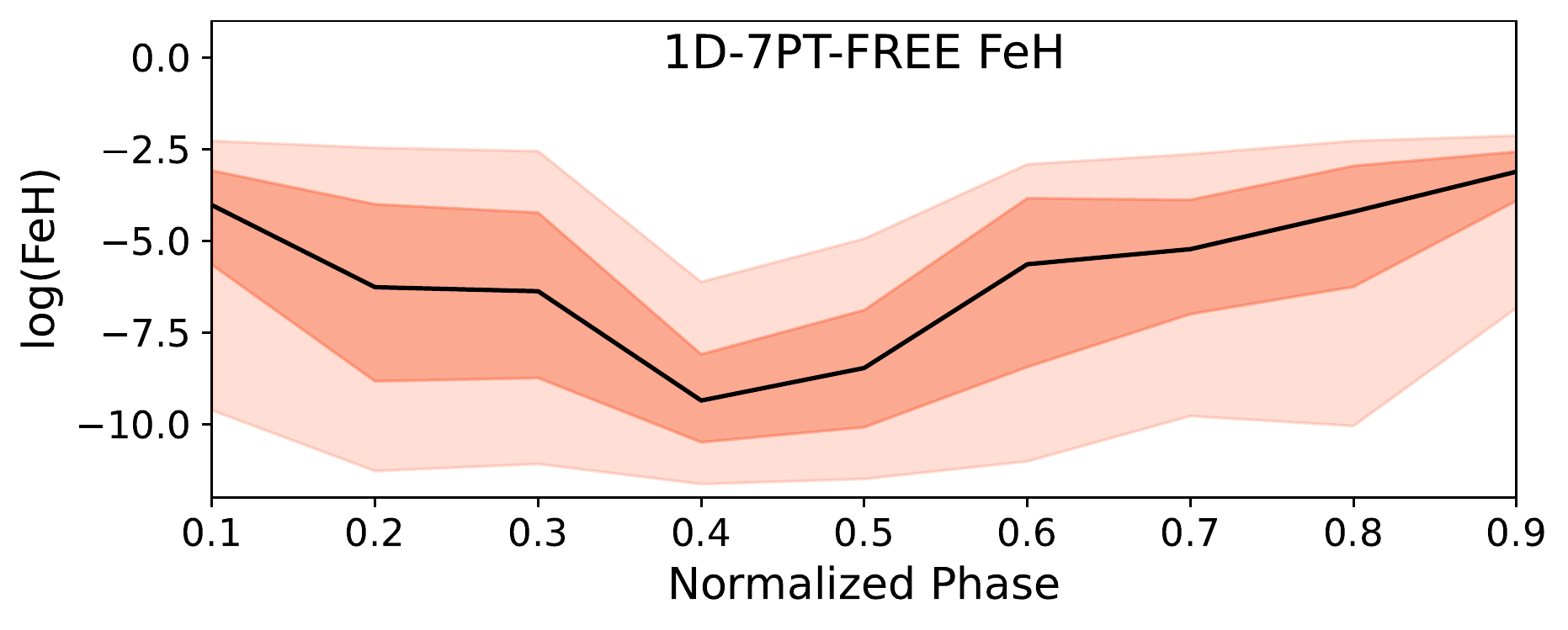}
    \includegraphics[width = 0.46\textwidth]{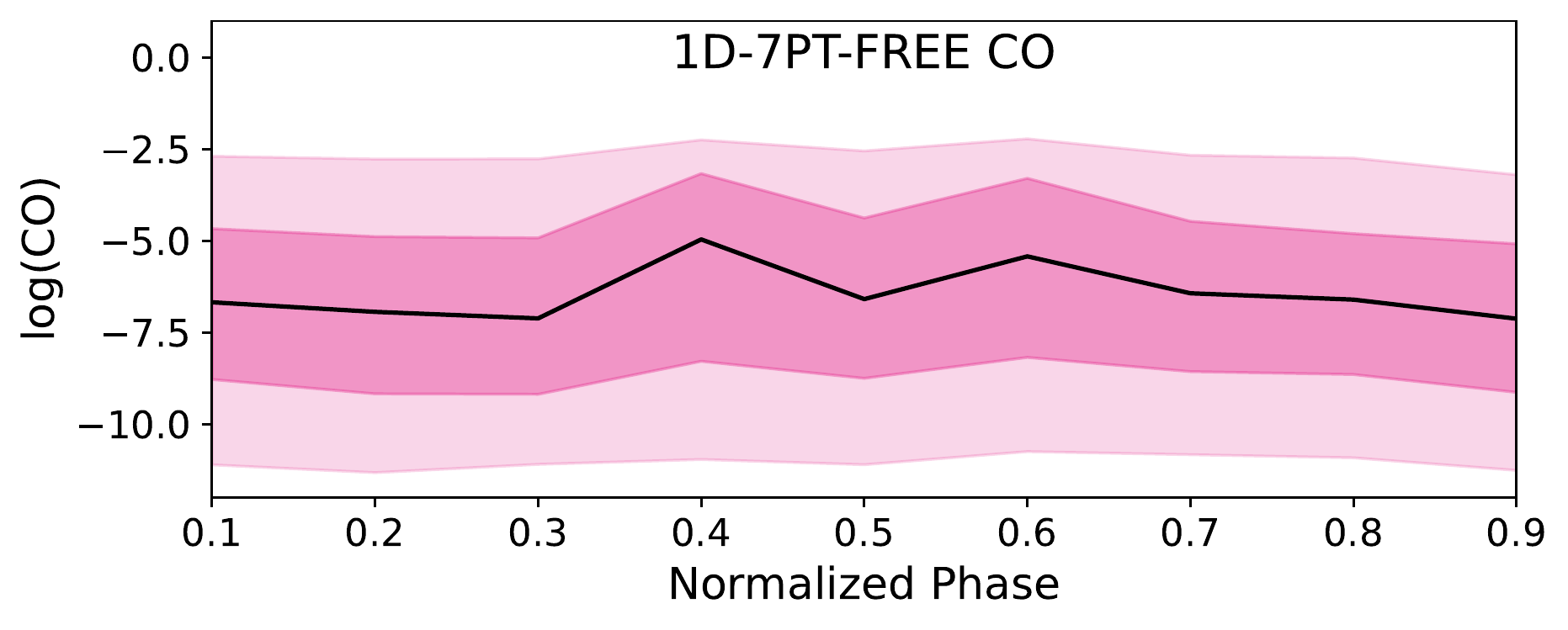}
    \includegraphics[width = 0.46\textwidth]{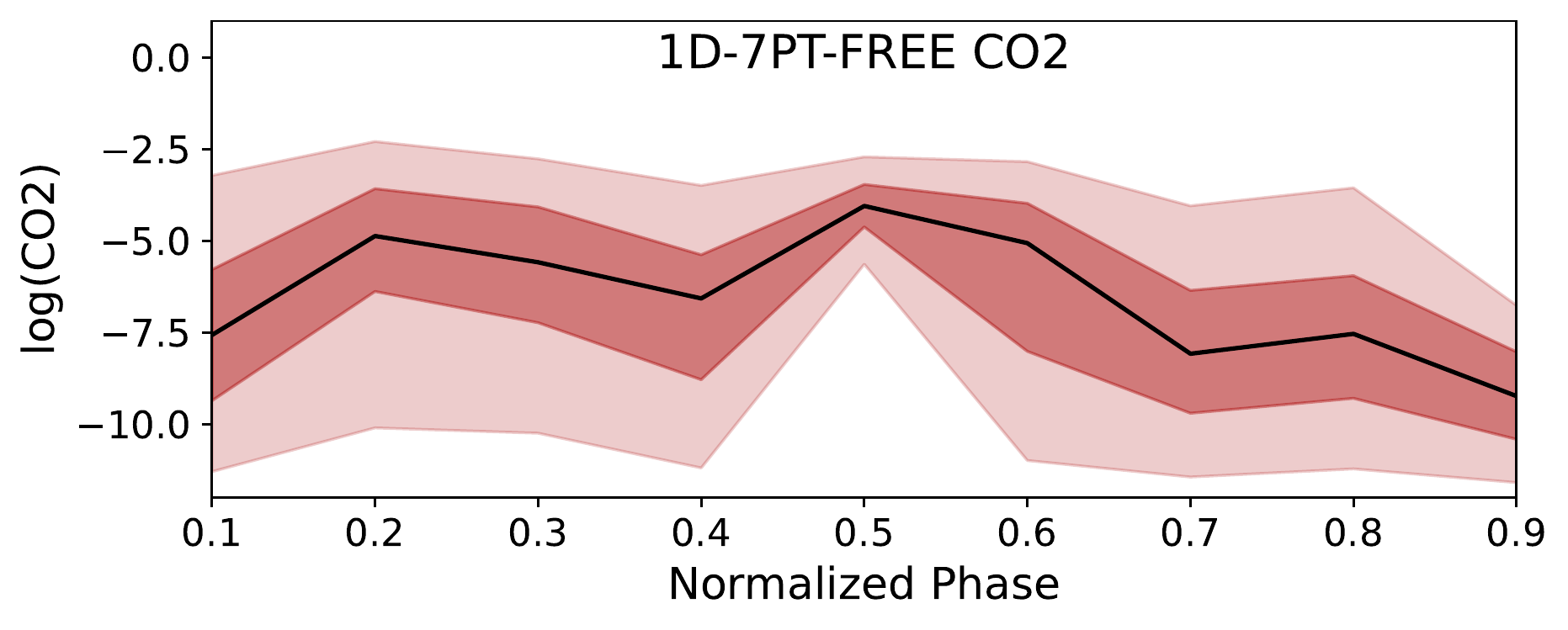}
    \includegraphics[width = 0.46\textwidth]{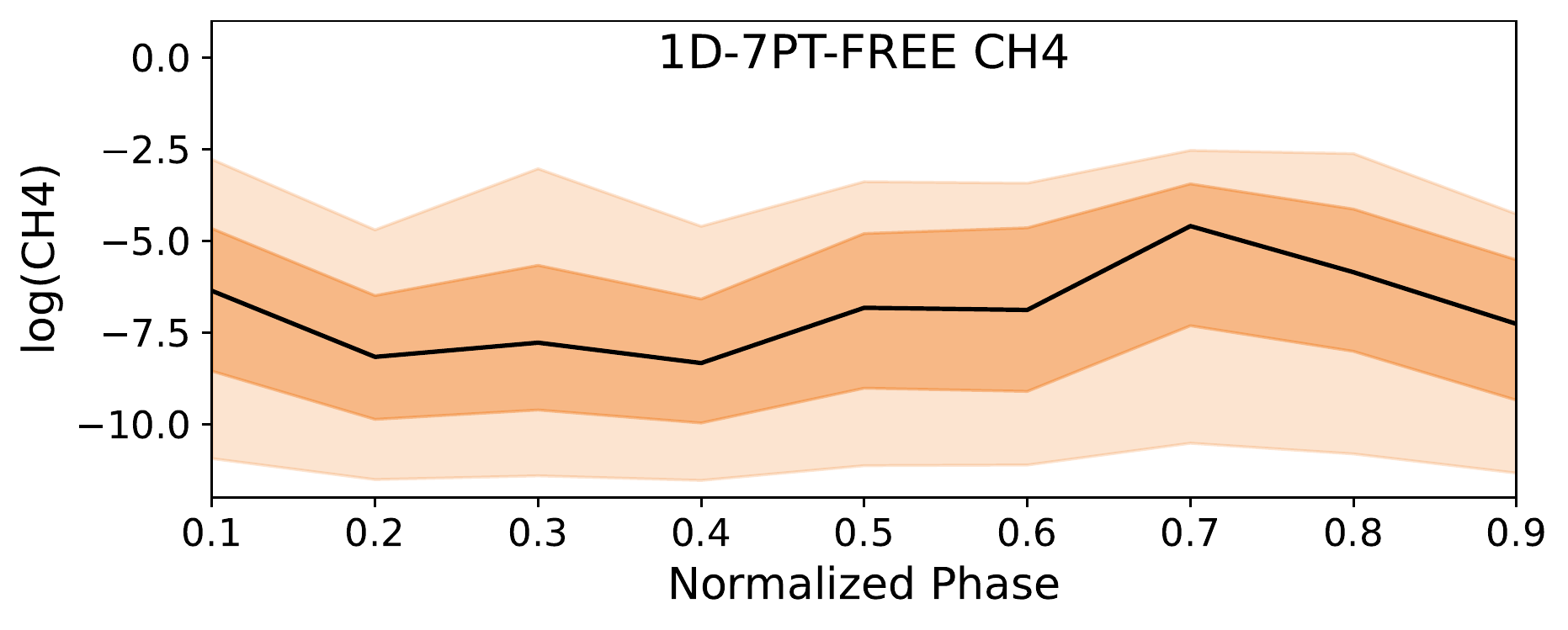}
    
    \caption{Retrieved abundances from the 1D-7PT-FREE retrievals. First column in descending order: TiO, FeH and CO$_2$; Second column: VO, CO and CH$_4$. Shaded regions represent 1$\sigma$ and 3$\sigma$ uncertainties.}\label{fig:tio_feh_map_free}
\end{figure}

Figure \ref{fig:tio_feh_map_free} shows the retrieved abundances depending on the phase for the main for the main molecules in the 1D-7PT-FREE retrievals. 

\subsection*{Appendix 4: Summary of the Bayesian evidence obtained in the 1D retrievals.}

Table \ref{tab:bayesian} indicates the Bayesian evidences and black-body temperatures obtained at each phases for the 1D runs.

\begin{table}[h]
\centering
\begin{tabular}{|l|l|l|l|l|l|l|l|l|l|}
\hline
Normalized Phase   & 0.1        & 0.2        & 0.3        & 0.4        & 0.5       & 0.6        & 0.7        & 0.8        & 0.9        \\ \hline
Temperature (K)        & 1831$^{+33}_{-51}$ & 2255$^{+24}_{-20}$ & 2631$^{+17}_{-17}$ & 2871$^{+11}_{-12}$ & 2976$^{+7}_{-8}$  & 2849$^{+11}_{-11}$ & 2601$^{+16}_{-16}$ & 2250$^{+24}_{-19}$ & 1884$^{+42}_{-39}$ \\ \hline
log(E)$_{blackbody}$       & 87.7$\pm$0.1  & 82.6$\pm$0.1  & 80.7$\pm$0.1  & 82.1$\pm$0.1  & 67.1$\pm$0.1 & 82.2$\pm$0.1  & 88.2$\pm$0.1  & 82.3$\pm$0.1  & 72.1$\pm$0.1  \\ \hline
log(E)$_{7PT-FREE}$ & 89.7$\pm$0.1  & 87.2$\pm$0.1  & 84.6$\pm$0.1  & 90.1$\pm$0.1  & 86.6$\pm$0.2 & 87.0$\pm$0.1  & 87.8$\pm$0.1  & 88.3$\pm$0.1  & 89.2$\pm$0.1  \\ \hline
log(E)$_{3PT-FREE}$ & 90.0$\pm$0.1  & 86.0$\pm$0.1  & 82.8$\pm$0.1  & 90.0$\pm$0.1  & 85.7$\pm$0.2 & 87.9$\pm$0.1  & 88.4$\pm$0.1  & 88.1$\pm$0.1  & 89.4$\pm$0.1  \\ \hline
log(E)$_{7PT-EQ}$   & 90.0$\pm$0.1  & 84.8$\pm$0.1  & 82.4$\pm$0.1  & 91.8$\pm$0.1  & 89.9$\pm$0.1 & 88.1$\pm$0.1  & 86.9$\pm$0.1  & 87.2$\pm$0.1  & 84.8$\pm$0.1  \\ \hline
log(E)$_{3PT-EQ}$   & 89.9$\pm$0.1  & 84.9$\pm$0.1  & 80.7$\pm$0.1  & 90.5$\pm$0.1  & 89.9$\pm$0.1 & 86.9$\pm$0.1  & 86.7$\pm$0.1  & 87.3$\pm$0.1  & 86.0$\pm$0.1  \\ \hline
\end{tabular}
\caption{Blackbody temperature and log Bayesian Evidence, log(E), obtained for the 1D retrievals. }\label{tab:bayesian}
\end{table}

\clearpage

\subsection*{Appendix 5: Complementary retrieved chemical profiles for the 1.5D-7PT-EQ retrieval}

\begin{figure}[h]
    \centering
    \includegraphics[width = 0.329\textwidth]{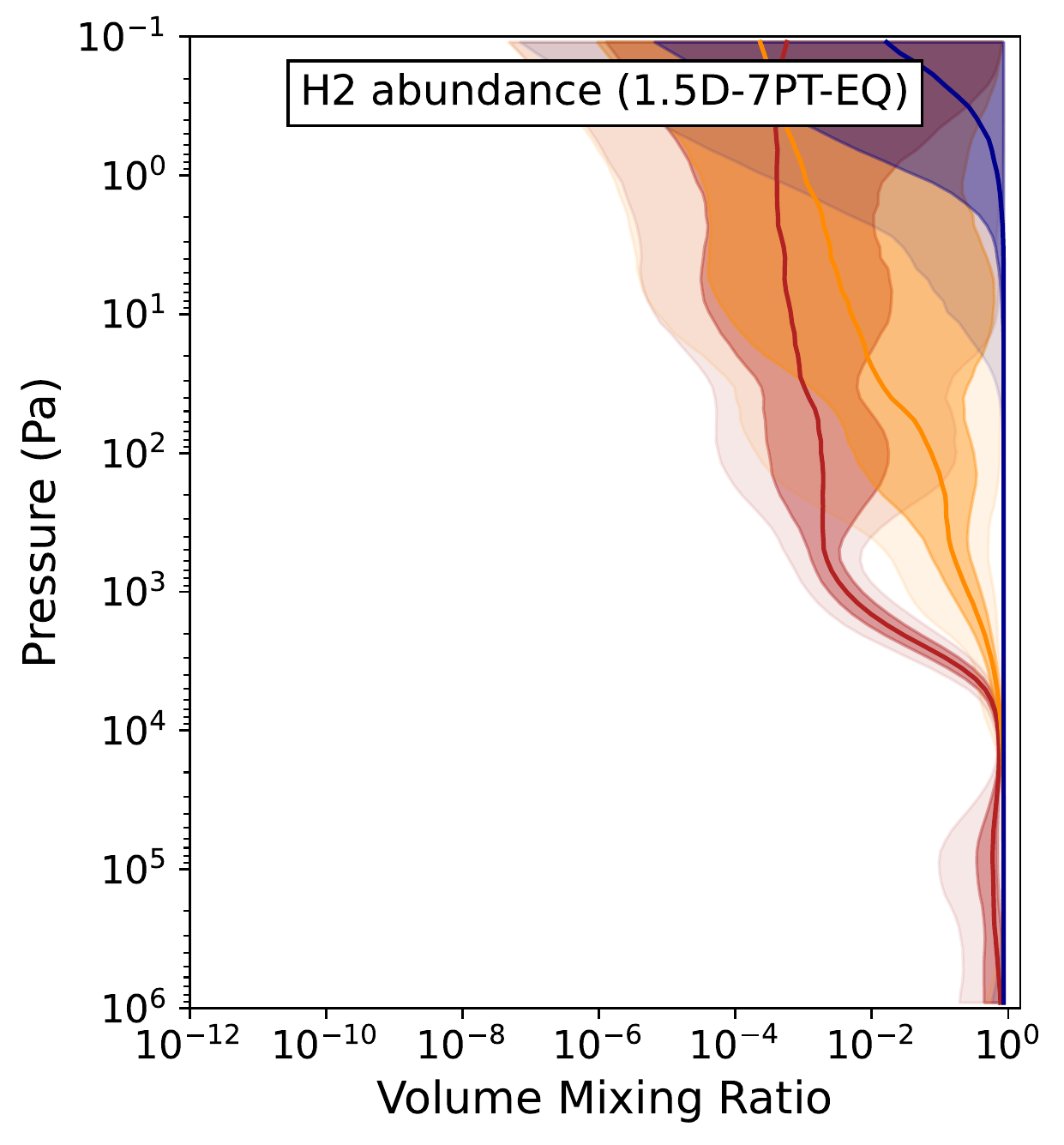}
    \includegraphics[width = 0.329\textwidth]{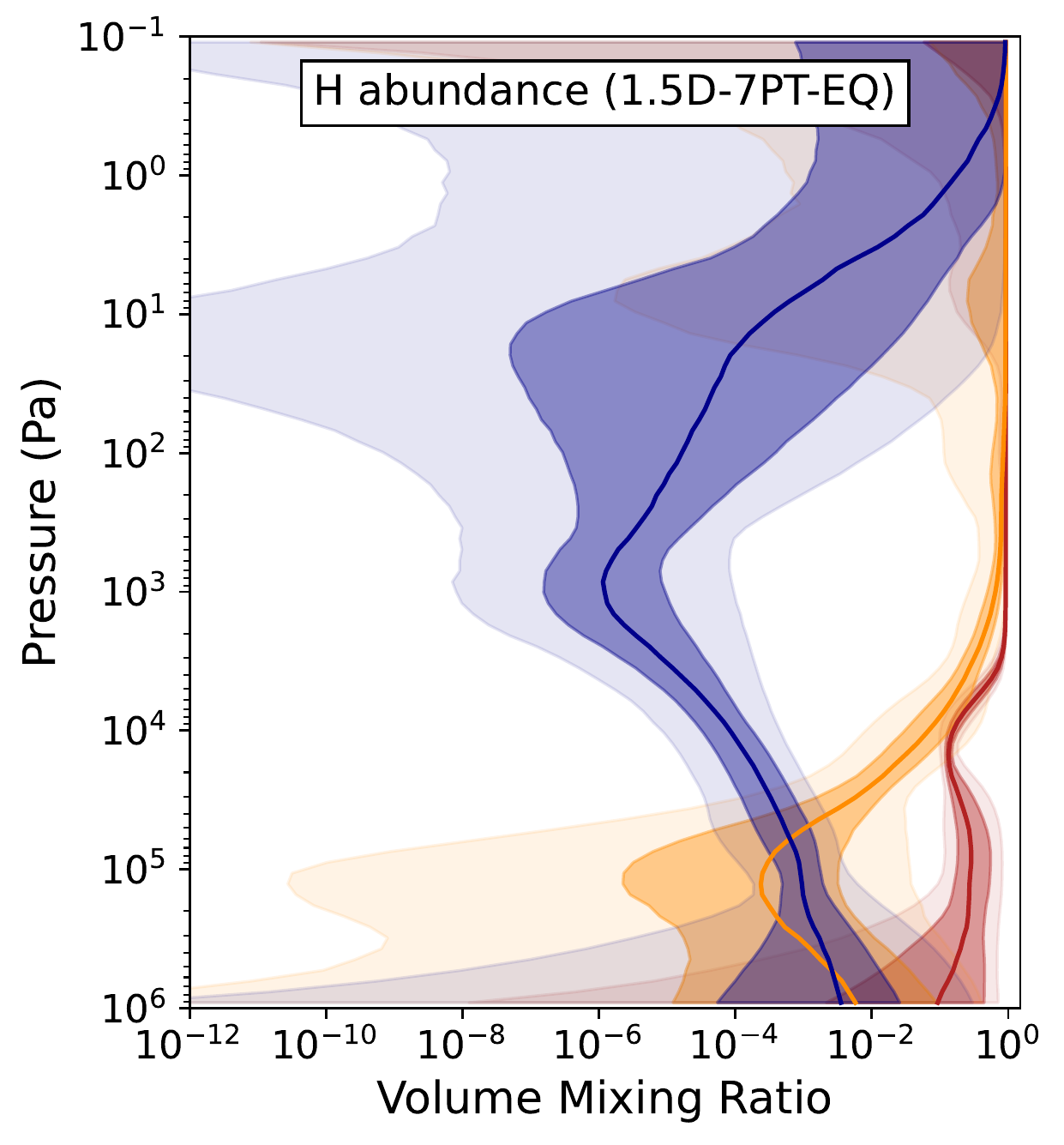}
    \includegraphics[width = 0.329\textwidth]{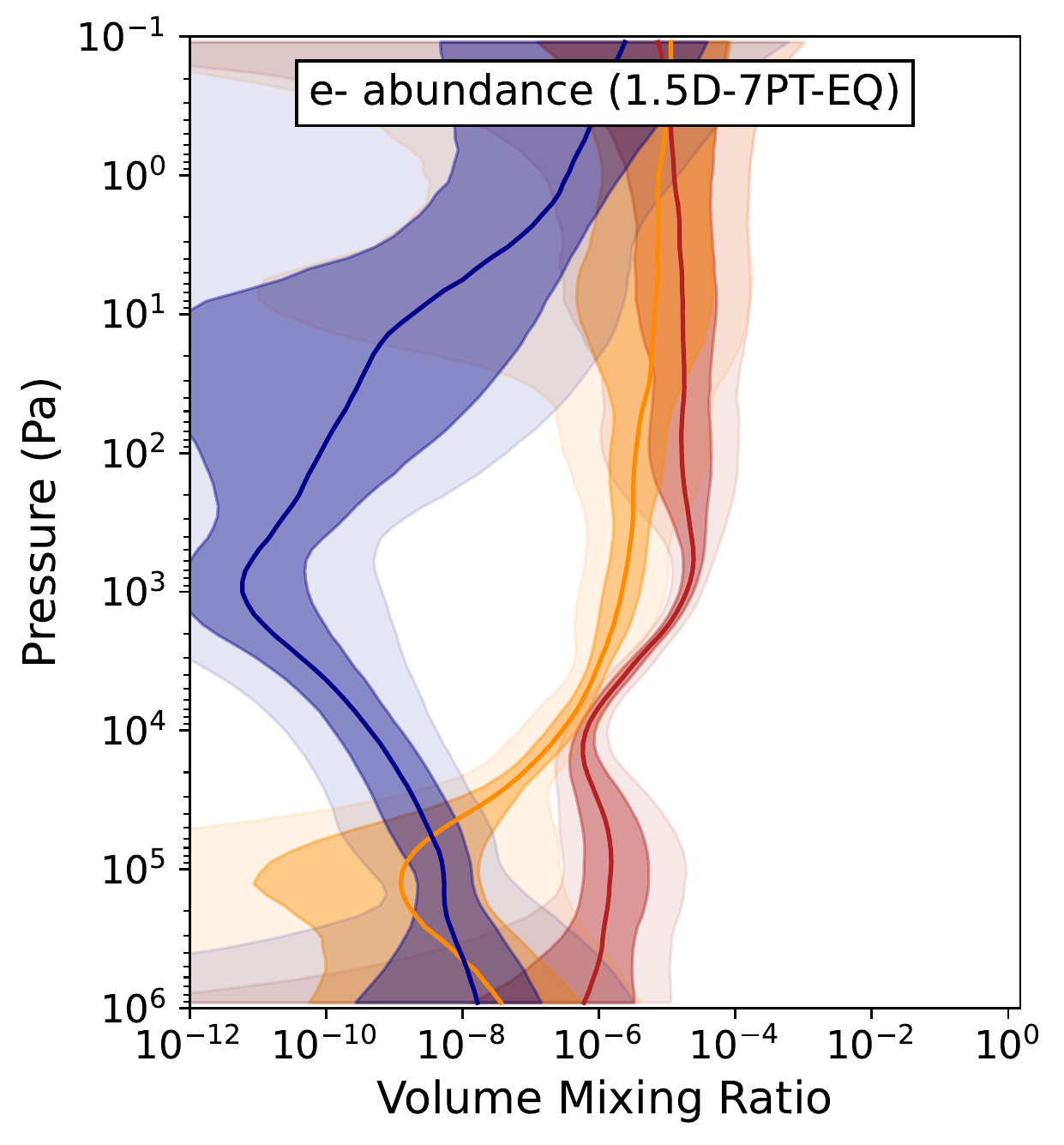}
    \includegraphics[width = 0.329\textwidth]{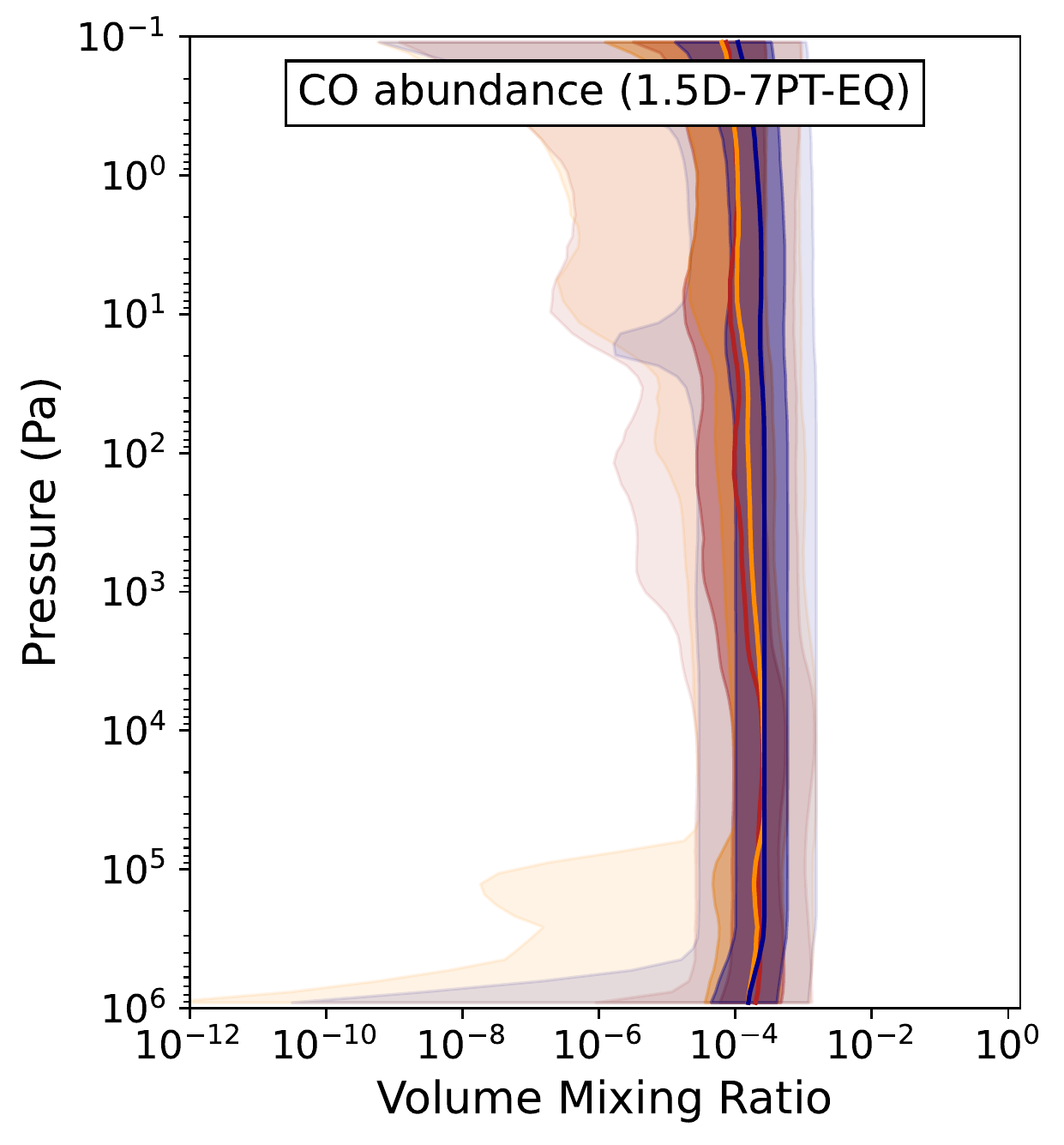}
    \includegraphics[width = 0.329\textwidth]{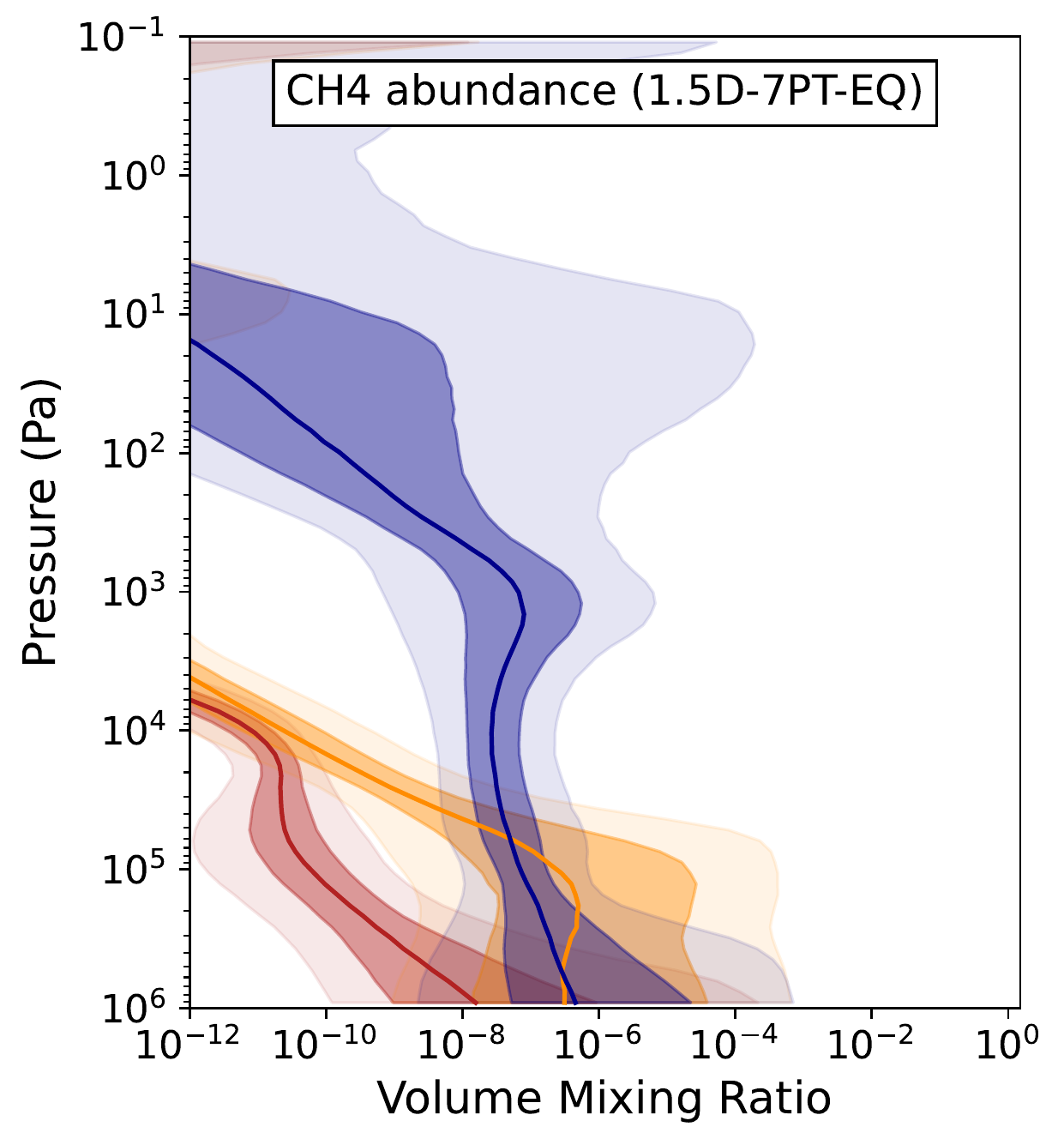}
    \includegraphics[width = 0.329\textwidth]{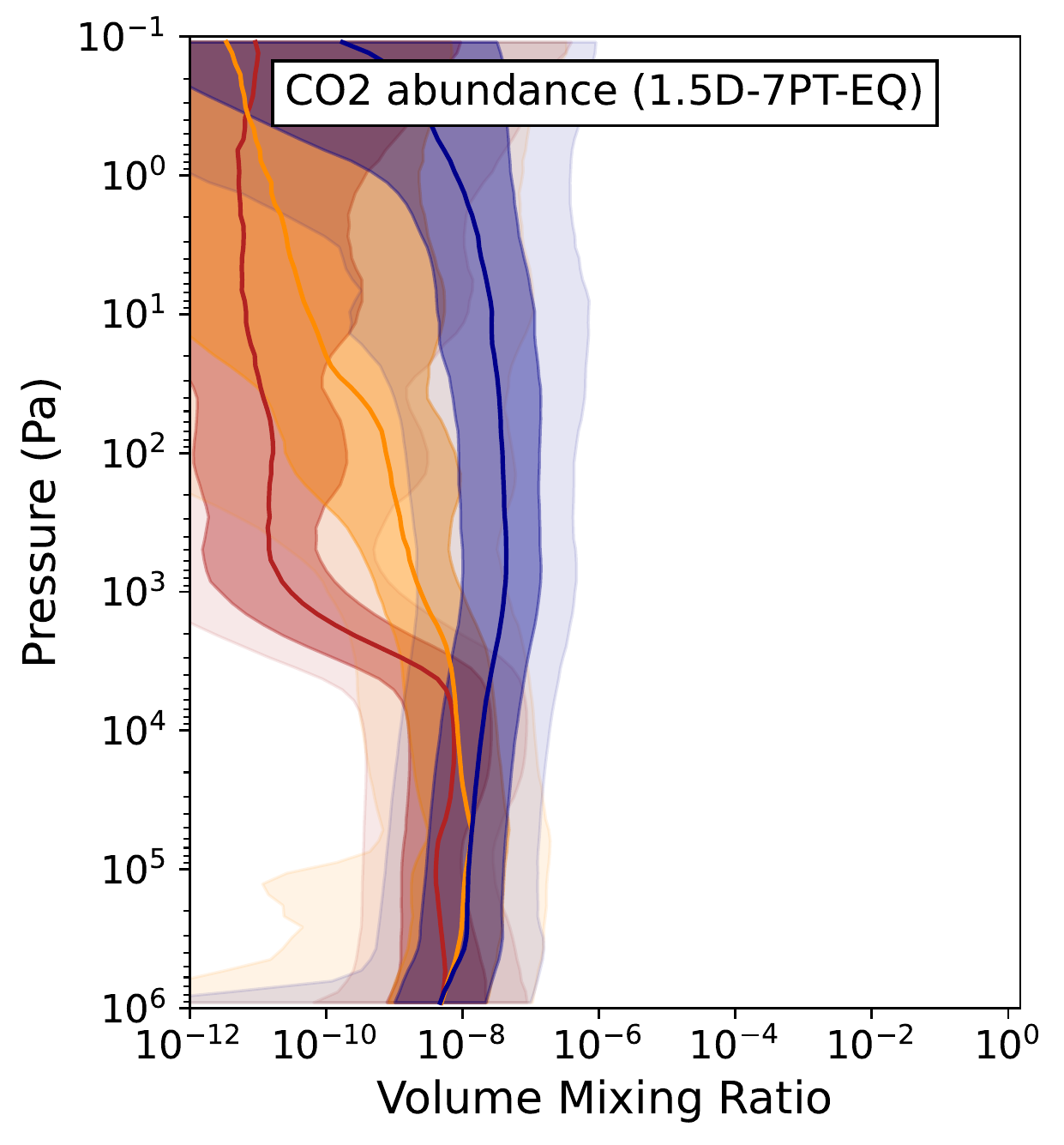}
    \includegraphics[width = 0.329\textwidth]{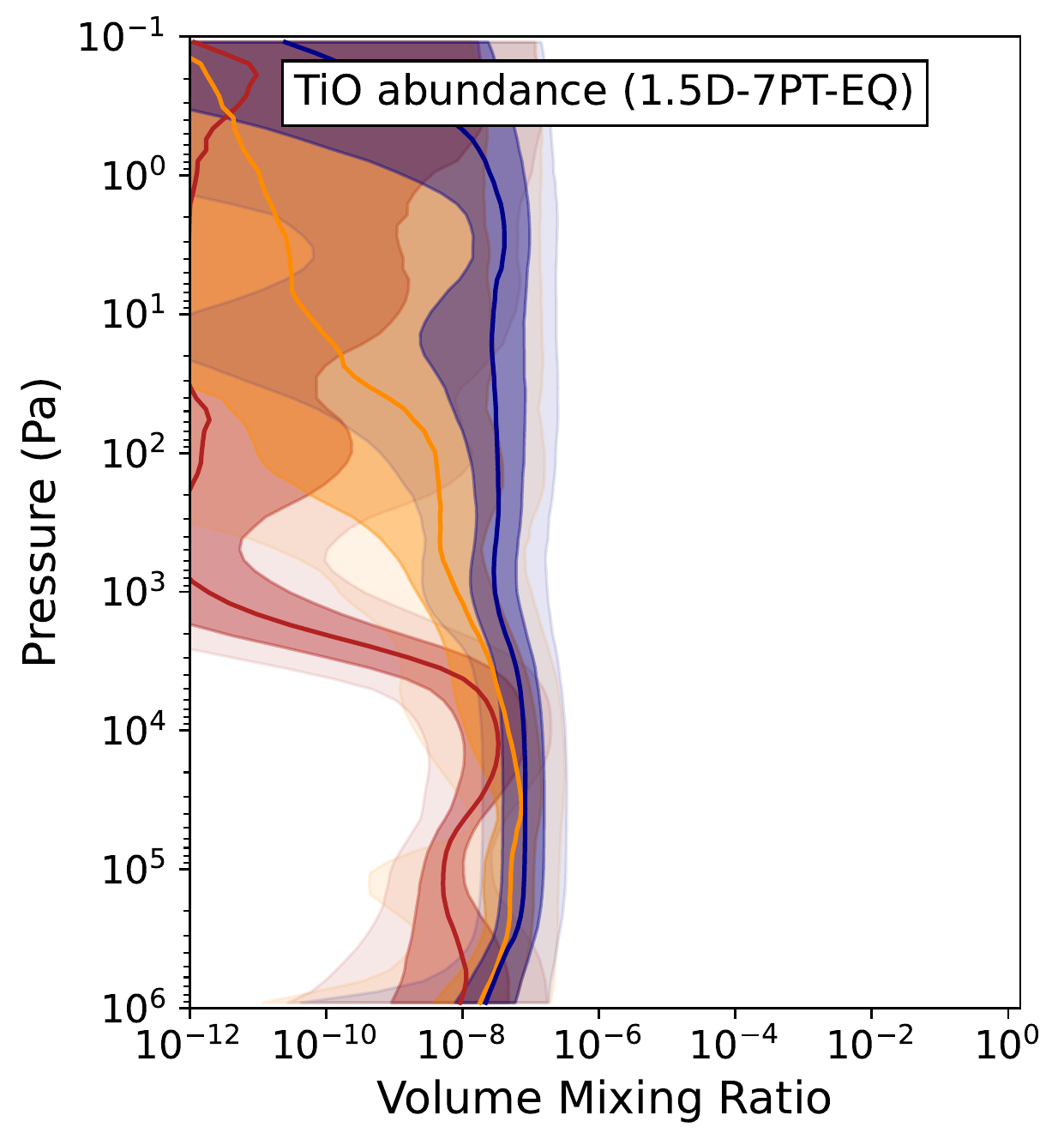}
    \includegraphics[width = 0.329\textwidth]{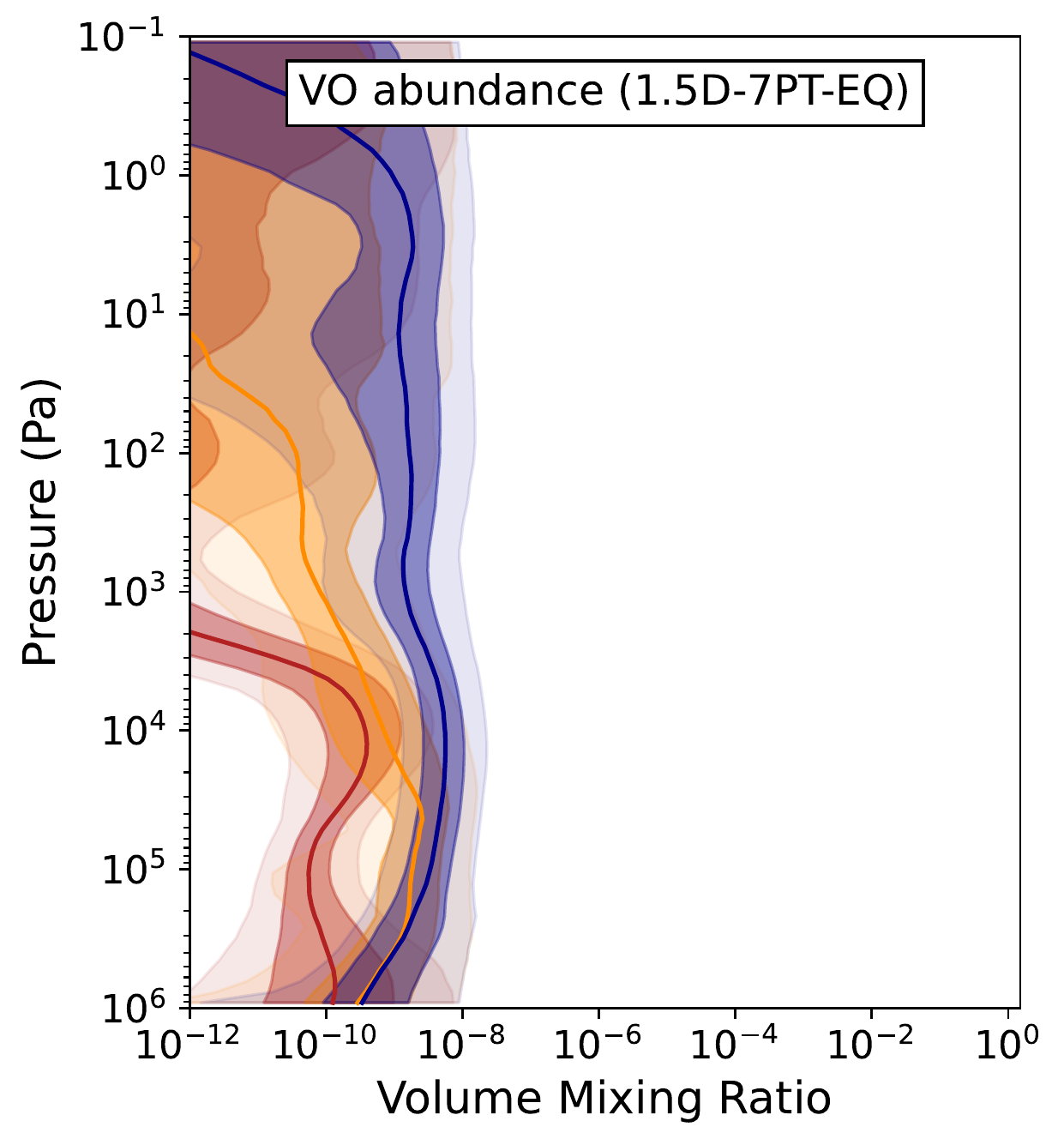}
    \includegraphics[width = 0.329\textwidth]{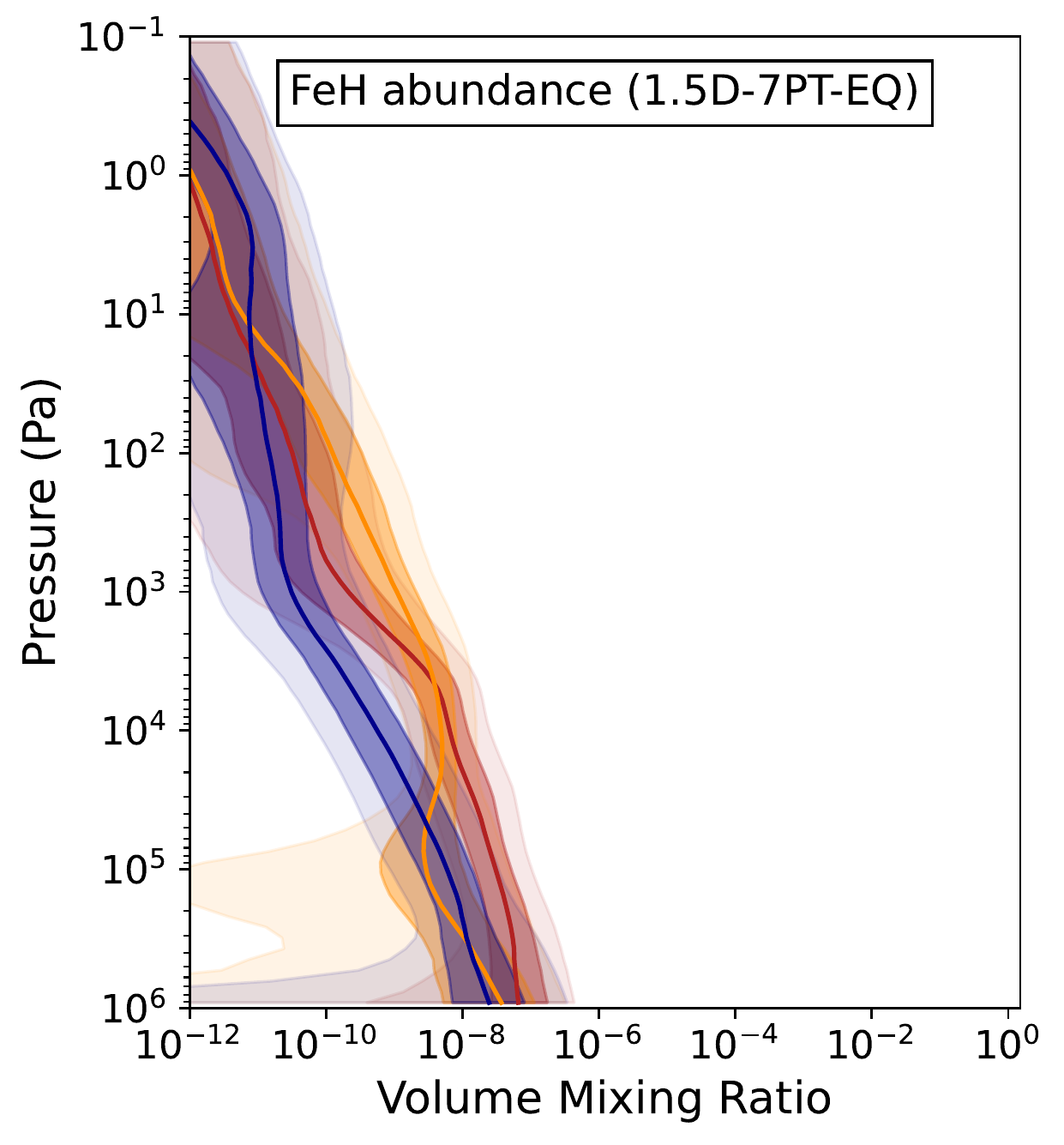}
    \caption{Retrieved volume mixing ratio of species in the 1.5D-7PT-EQ retrieval. Red: hot-spot; Orange: day-side; Blue: night-side.}
    \label{fig:1.5d_7pt_eq_chem}
\end{figure}

Figure \ref{fig:1.5d_7pt_eq_chem} shows the abundance profiles obtained in the 1.5D-7PT-EQ retrieval.

\clearpage

\subsection*{Appendix 6: Full posterior distribution of the 1.5D-7PT-FREE retrieval}

\begin{figure}[h]
    \centering
    \includegraphics[width = 0.99\textwidth]{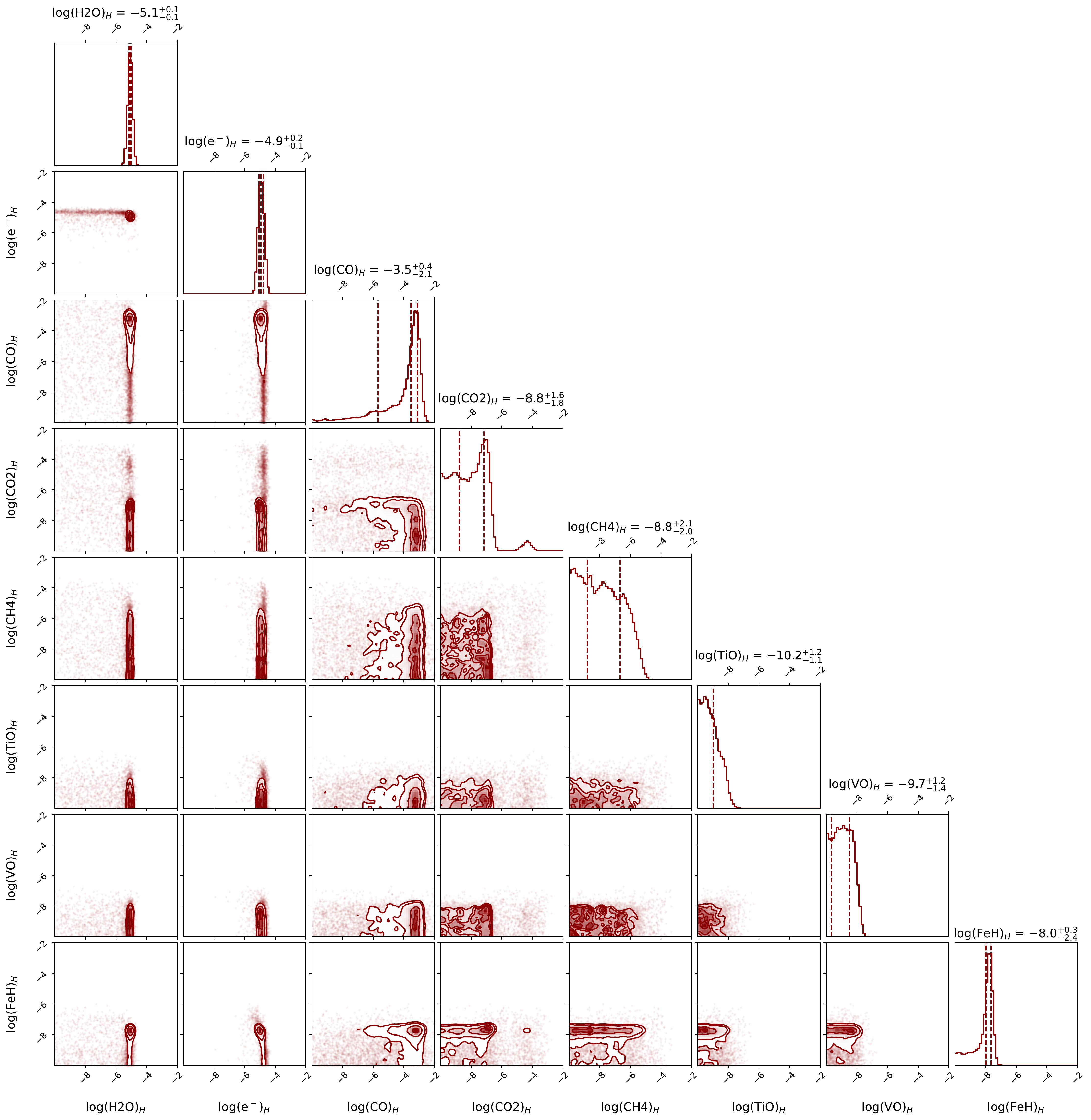}
    \caption{Full posterior distribution of the 1.5D-7PT-FREE retrieval for the hot-spot. The Temperature-Pressure profile was fixed to the mean of the 1.5D-7PT-EQ run.}
    \label{fig:1.5d_7pt_free_post_hs}
\end{figure}

\begin{figure}
    \centering
    \includegraphics[width = 0.99\textwidth]{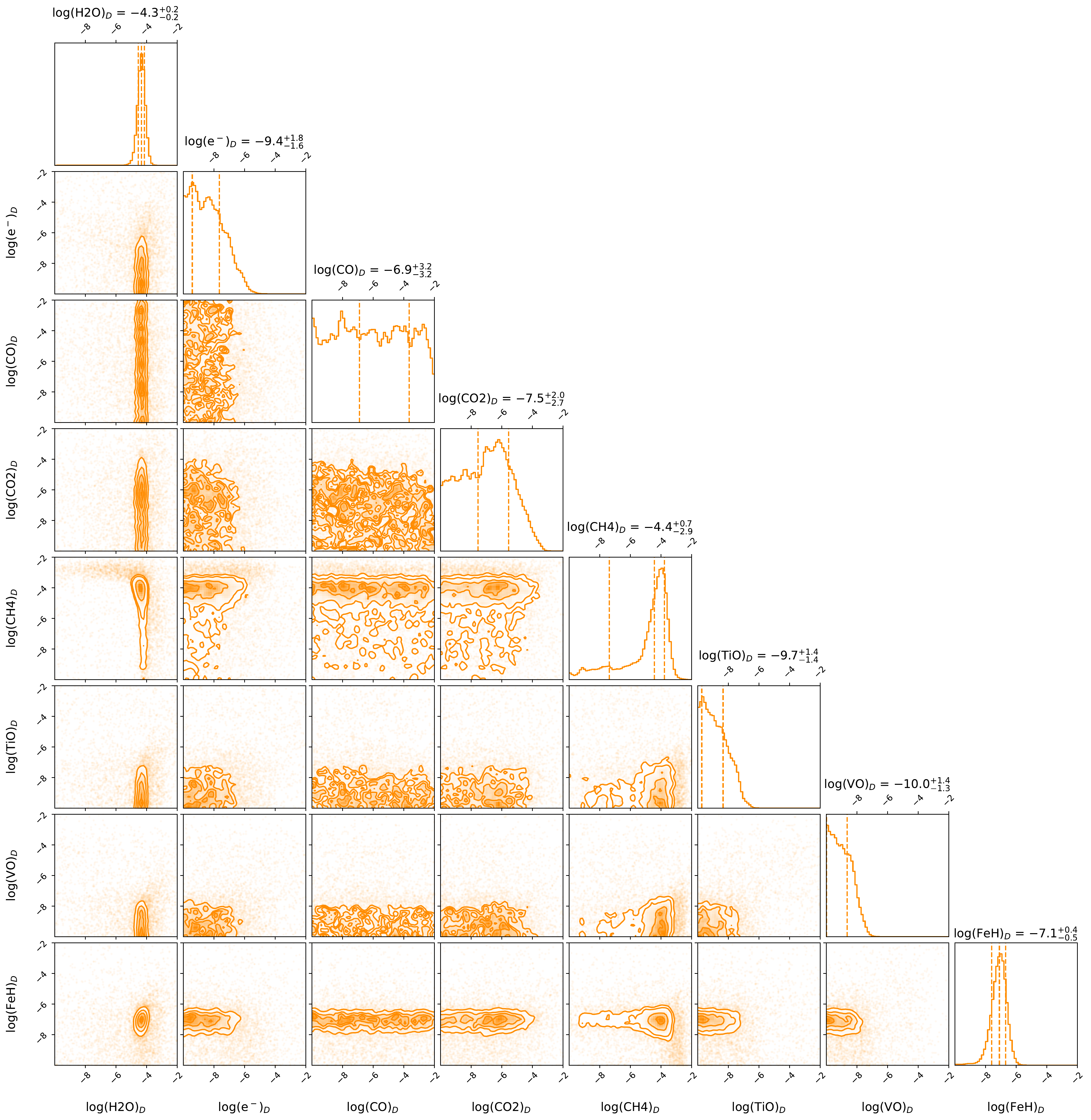}
    \caption{Full posterior distribution of the 1.5D-7PT-FREE retrieval for the day-side. The Temperature-Pressure profile was fixed to the mean of the 1.5D-7PT-EQ run.}
    \label{fig:1.5d_7pt_free_post_day}
\end{figure}

\begin{figure}
    \centering
    \includegraphics[width = 0.99\textwidth]{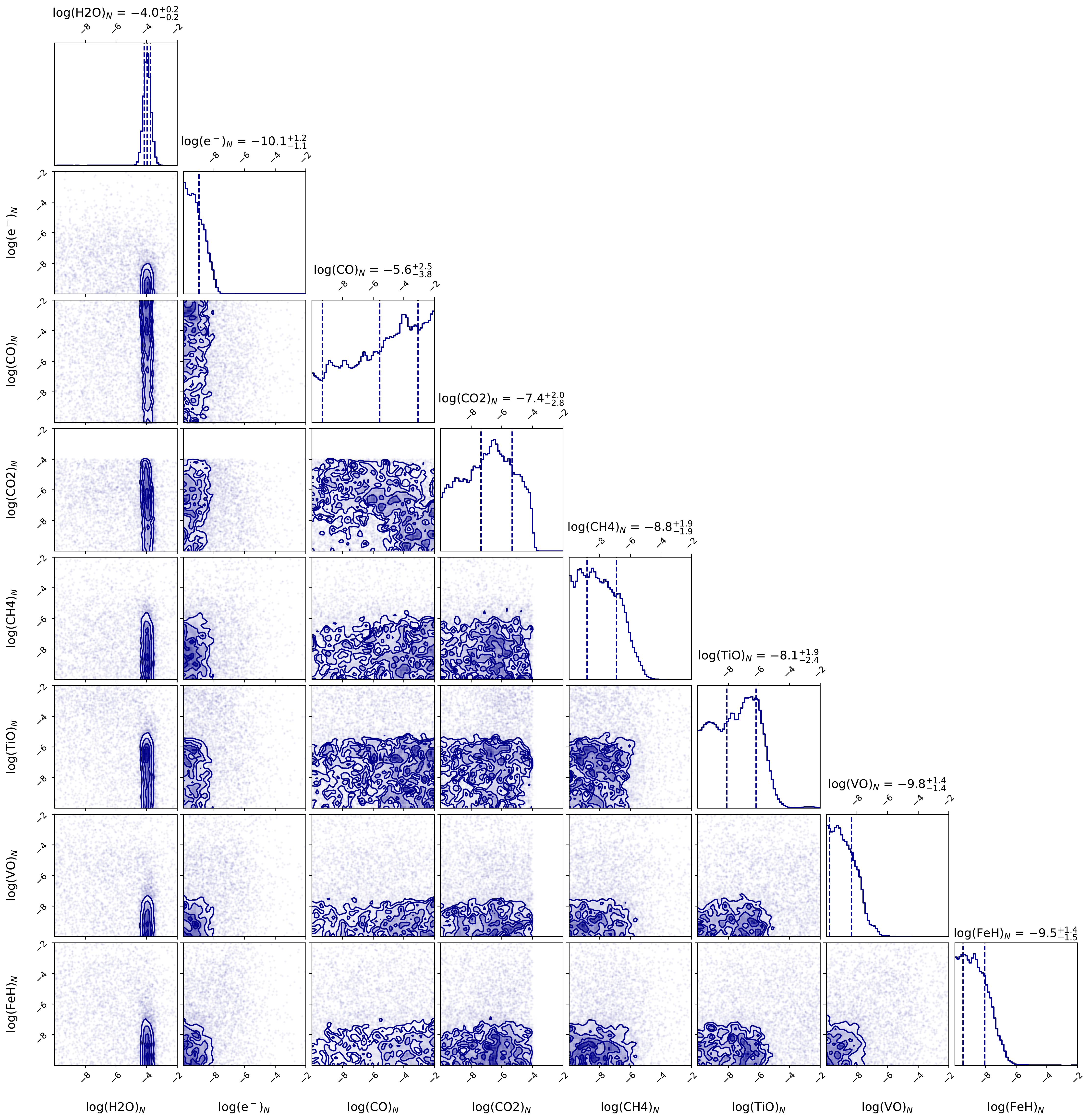}
    \caption{Full posterior distribution of the 1.5D-7PT-FREE retrieval for the night-side. The Temperature-Pressure profile was fixed to the mean of the 1.5D-7PT-EQ run.}
    \label{fig:1.5d_7pt_free_post_night}
\end{figure}

The posterior distributions of the 1.5D-7PT-FREE retrieval are shown in Figure \ref{fig:1.5d_7pt_free_post_hs} for the hot-spot, Figure \ref{fig:1.5d_7pt_free_post_day} for the day-side and Figure \ref{fig:1.5d_7pt_free_post_night} for the night side.

\clearpage

\subsection*{Appendix 7: Animated spectra and thermal profiles for the 1D and 1.5D phase-curve retrievals performed in this study}

\begin{figure}[h]
\centering
\begin{interactive}{animation}{ANIMATION_1D_V3_DARK_syn1.mp4}
\includegraphics[width = 0.90\textwidth]{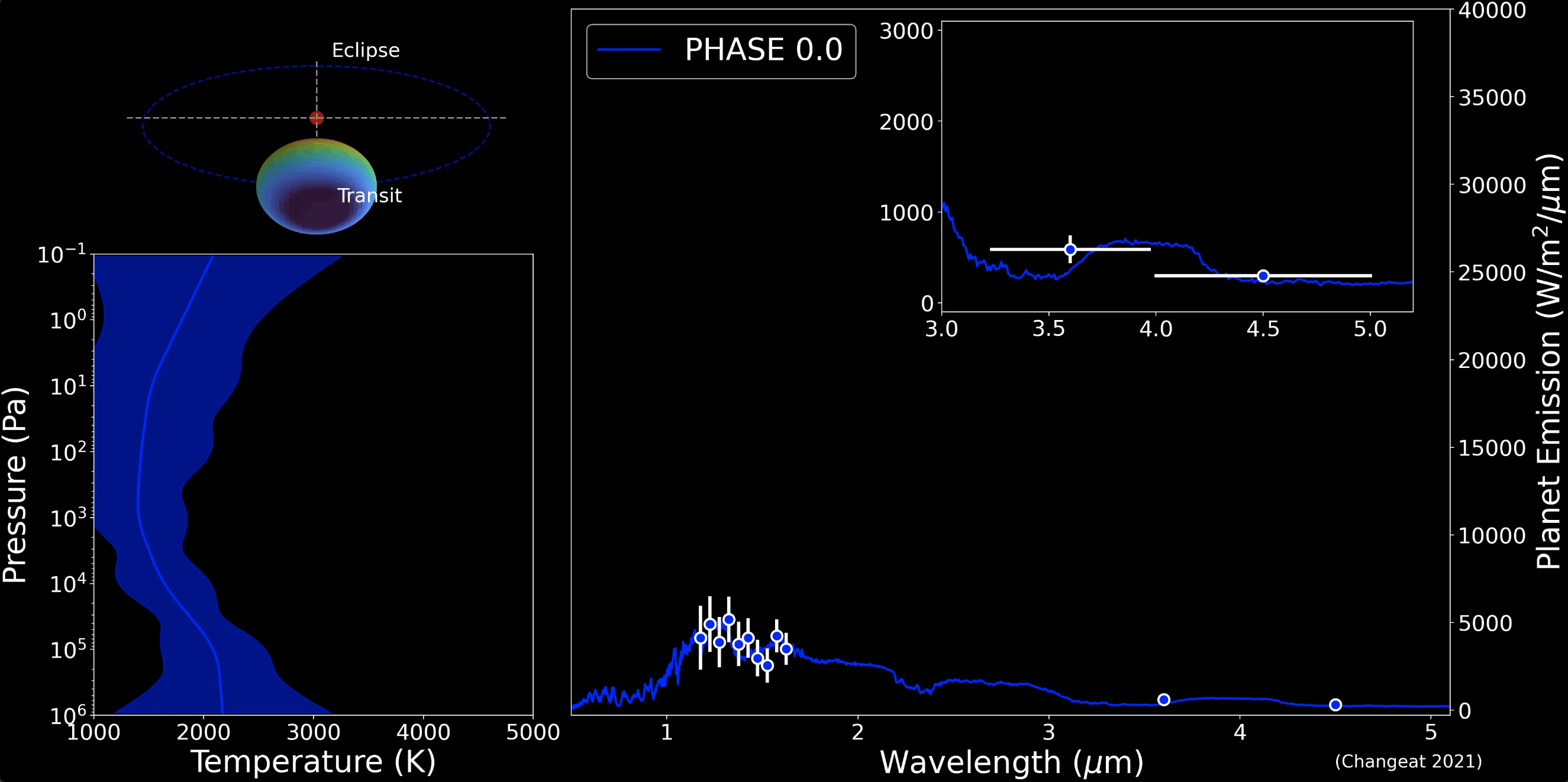}
\end{interactive}

\begin{interactive}{animation}{ANIMATION_15D_V3_DARK_syn1.mp4}
\includegraphics[width = 0.90\textwidth]{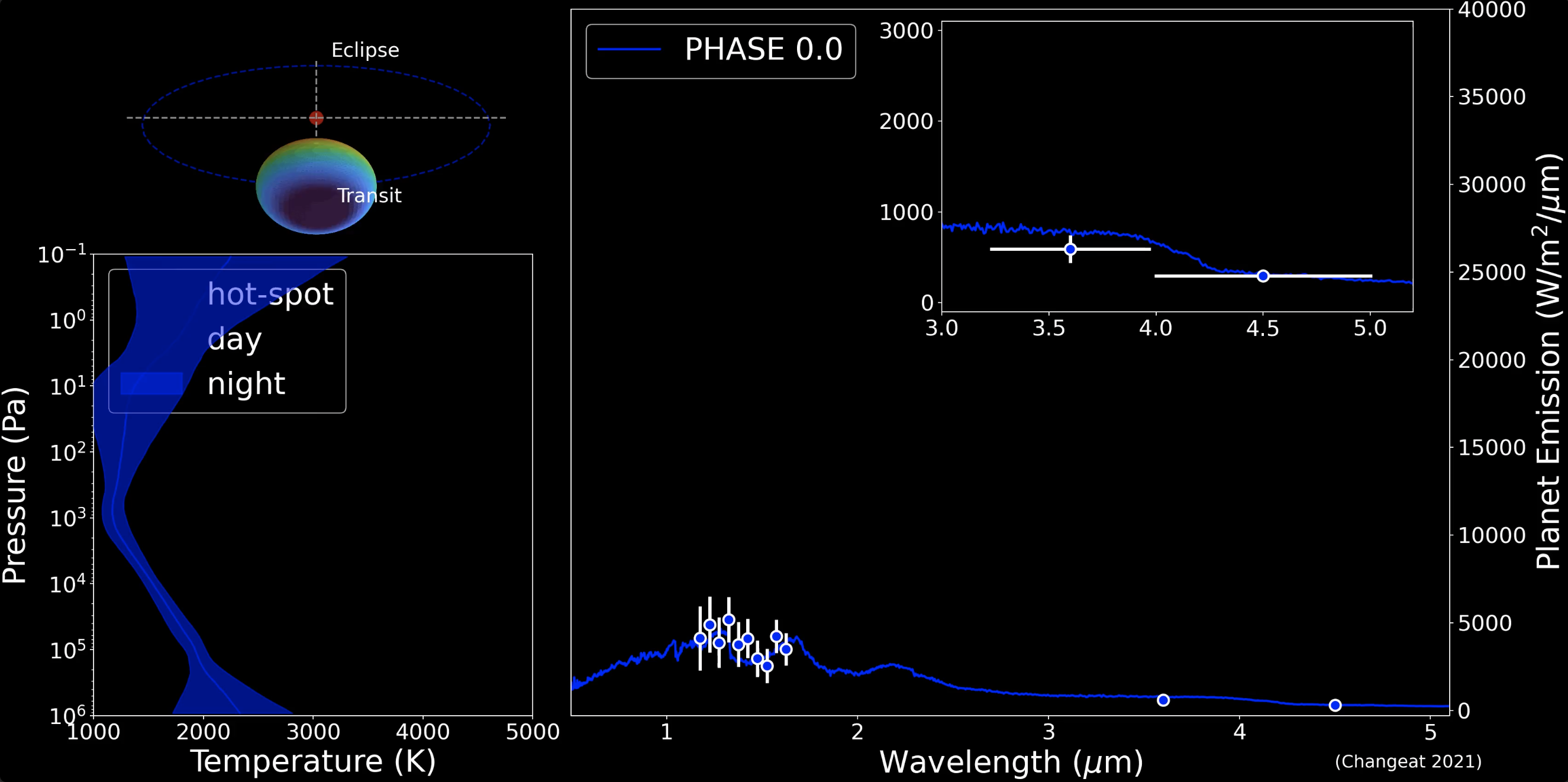}
\end{interactive}
\caption{Spectra and thermal profiles for the 1D-7PT-EQ (top) and the 1.5D-7PT-EQ (bottom) retrievals at phase 0. An animated version of this figure provides the same information but evolving with time. In the 1D-7PT-EQ case, a 1D retrieval is performed for each analysed phase. On the contrary, in the 1.5D retrieval all the phases are analysed at once. The transparency of the thermal profiles in the 1.5D retrieval shows the contribution of each region to the different observed phase.} \label{fig:retrieval_animation}
\end{figure}

Figure \ref{fig:retrieval_animation} shows an animation of the best fit spectra and thermal structure for the 1D-7PT-EQ and the 1.5D-7PT-EQ cases.

\end{document}